\documentclass[11pt]{article}

\usepackage{amsmath,amssymb}
\usepackage{slashed}
\usepackage{xcolor} 
\usepackage{graphicx}
\usepackage{dcolumn} 
\usepackage{bm} 
\usepackage{epsfig}
\usepackage{epstopdf}
\usepackage{grffile}
\usepackage{color}
\usepackage{colordvi}
\usepackage{amsmath,amssymb}
\usepackage{rotating}
\usepackage{lscape}
\usepackage{cite}
\usepackage{float}
\usepackage{hyperref}
\usepackage[utf8]{inputenc}

\usepackage{lineno}

\usepackage{physics}
\usepackage{graphicx,bm}

\def\Re{{\cal R \mskip-4mu \lower.1ex \hbox{\it e}\,}}
\def\Im{{\cal I \mskip-5mu \lower.1ex \hbox{\it m}\,}}

\def\tev{\,{\ifmmode\mathrm {TeV}\else TeV\fi}}
\def\gev{\,{\ifmmode\mathrm {GeV}\else GeV\fi}}
\def\mev{\,{\ifmmode\mathrm {MeV}\else MeV\fi}}
\def\to{\rightarrow}

\newcommand{\fb}{\text{fb}}

\newcommand{\beq}{\begin{equation}}
\newcommand{\bea}{\begin{eqnarray}}
\newcommand{\eeq}{\end{equation}}
\newcommand{\eea}{\end{eqnarray}}
\newcommand{\bal}{\begin{align}}
\newcommand{\eal}{\end{align}}

\usepackage{tikz}
\usetikzlibrary{decorations.pathmorphing,decorations.markings}
\tikzset{
photon/.style={decorate, decoration={snake,amplitude=4pt, segment length=7pt}, draw=black},
particle/.style={draw=black, postaction={decorate}, decoration={markings,mark=at position .5 with {\arrow[draw=black]{>}}}},
antiparticle/.style={draw=black, postaction={decorate}, decoration={markings,mark=at position .5 with {\arrow[draw=black]{<}}}},
gluon/.style={decorate, draw=black, decoration={coil,amplitude=3pt, segment length=4pt}},
higgs/.style={draw=black,dashed,thick },
arrow/.style={draw=black, very thick, postaction={decorate}, decoration={markings,mark=at position 1 with {\arrow[draw=black]{>}}}}
}

\usepackage{latexsym}
\usepackage{subcaption}

\usepackage{outlines}
\usepackage{enumitem}
\setenumerate[1]{label=\Roman*.}
\setenumerate[2]{label=\Alph*.}
\setenumerate[3]{label=\roman*.}
\setenumerate[4]{label=\alph*.}

\definecolor{darklightsabergreen}{rgb}{0.0, .49, 0.06}

\begin{document}

\begin{center}

\vspace*{15mm}
\vspace{1cm}
{\Large \bf  Same-sign top pair (plus a $W$) production in flavor changing vector and scalar models }

\vspace{1cm}

{\bf Javad Ebadi$^{\dagger}$,  Fatemeh Elahi$^{\ddagger}$,  Morteza Khatiri$^{\dagger}$,  Mojtaba Mohammadi Najafabadi$^{\ddagger}$ }

 \vspace*{0.5cm}

{\small\sl 
$^{\dagger}$School of Physics, Institute for Research in Fundamental Sciences, P.O. Box 19395-5531,  Tehran, Iran  \\
$^{\ddagger}$School of Particles and Accelerators, Institute for Research in Fundamental Sciences (IPM) P.O. Box 19395-5531, Tehran, Iran }\\

\vspace*{.2cm}
\end{center}

\vspace*{0.5cm}
\begin{abstract}
\vspace*{0.5cm}

{
We investigate the prospect of the LHC for discovering new physics effects
via new strategies in the same-sign top pair and same-sign top pair associated with a $W$ boson signatures.
Significant enhancement in production of same-sign top quarks (plus a $W$ boson) 
is a joint property of several models beyond the standard model.
We concentrate on the leptonic (electron and muon) decay of the top quarks
and study the exclusion reach of the LHC data for a simplified model approach 
where top quark flavor changing could occur through a $Z'$ or a neutral scalar $\phi$ exchange.
Less background contributions and clean signature are the advantages of the leptonic decay 
mode of the top quarks in the same-sign production processes.
A combination is performed on both same-sign top pair and same-sign top pair
plus a $W$ boson production modes which enables us to reach a large fraction
of the model parameter space. Assuming the couplings of new physics of the order of $10^{-2}$, 
the mass of a flavor changing $Z'$ or a neutral scalar above  1 TeV could be excluded. 
We propose a momentum dependent charge asymmetry
and angular observables in the same-sign top process which provide the 
possibility of separation of new physics signal from the SM backgrounds
as well as discrimination of the flavor changing  $tuX$ from $tcX$, where $X = Z',\phi$.
} 
\end{abstract}
\newpage

\section{ Introduction}
\label{sec:Intro}

Despite the marvelous successes of the Standard Model (SM) at the electroweak scale,
we strongly believe there exists new physics (NP) beyond the SM.
Among all fermions in the SM, top quark causes the most serious 
hierarchy and plays an essential role in the vacuum stability \cite{Buttazzo:2013uya}.
Top quark has the strongest coupling with the Higgs boson
as a result the quantum loop level effects on precise measured
observables could be significant.  Such strong effects can be connected
to the couplings of new degrees of freedom to the top quark. 

Within the SM framework, flavor changing neutral currents (FCNCs) in the top quark sector
are not present at leading-order in both Yukawa and gauge 
interactions.  However, the FCNC couplings could
be generated from loop-level box  diagrams
which are extremely suppressed due to the Glashow-Iliopoulos-Maiani (GIM) mechanism \cite{Glashow:1970gm}.
We note that any FCNC interaction occurring between the first two
quark generations is significantly constrained by the low energy experiments \cite{eftfcnc}.
However, there is a large concentration on the top quark FCNC couplings as
the constraints are rather mild. 
The top quark FCNC processes could 
 raise considerably in the presence of well-motivated new physics models.
Among these models, there are scenarios such as supersymmetric extensions of the SM and the 
two Higgs doublet models where the FCNCs increase 
because of the new loop level diagrams
mediated by new particles
 \cite{ deDivitiis:1997sh , Cao:2007dk, Cao:2004wd , Lopez:1997xv , Guasch:1999jp,Liu:2004qw ,Agashe:2006wa ,Agashe:2009di, soni, thdm}. Moreover, there are  proposed theories beyond the SM
in which top quark FCNCs could show up through the exchange of a new neutral scalar or a neutral gauge boson.
Some of the models attempt to explain the flavor structure of the SM, where a new mediator 
 is introduced with couplings that are stronger to quarks with larger masses. 
 If the mediator is a scalar, which acquires a vacuum expectation value ($vev$)  similar to the Higgs mechanism, 
 its couplings to fermions is proportional to their masses \cite{ Bauer:2016rxs, Gaitan:2016rlq, Hall:1993ca, Azatov:2009na, Agashe:2009di,   Agram:2013wda, Atwood:2013xg,Alvarado:2017bax}.
In addition, the top quark FCNC could occur via a $Z'$ boson exchange which can originate from 
Grand Unified Theories (GUT).  In such models, 
a $Z'$ boson can couple universally to the three fermion generations in a flavor diagonal way
with a possibility of mixing all SM quarks by $Z$ and $Z'$ bosons \cite{Arhrib:2006sg}. The flavor changing 
$Z'$ has been also expected in theories with dynamical symmetry breaking approach. 
The top quark flavor changing have already been studied in many papers in the
context of non-universality, where the strength of the couplings are different among 
different generations \cite{Langacker:1991pg,Perez:1992hc, Arhrib:2006sg, Cakir:2010rs,
 Perez:2004jc, He:2007iu, He:2009ie, Gedalia:2010mf,Gupta:2010wt, Aranda:2009rj, Aranda:2009cd, Frank:2006ku, Aranda:2010cy}.	
There are also beyond the SM theories where the flavor changing could proceed through
color sextet vector and scalar bosons \cite{Zhang:2010kr,Berger:2010fy}.

Other class of beyond the SM theories 
leading to possible top quark FCNCs through a neutral scalar $\phi$
or a vector $Z'$ is related to the hidden sector. 
In these models, the dark matter candidate from the hidden sector 
can couple to the SM fields in a flavor changing way through a vector or a scalar mediator \cite{Andrea:2011ws, Kamenik:2011nb}.
Such models lead to production of a top quark associated
with dark matter or in association with a  $Z'$ that decays into
dark matter at the LHC \cite{Agram:2013wda, Boucheneb:2014wza}. These processes have a final state of
a top quark plus large missing energy (so called monotop) which is not present at leading-order
in the SM. Searches for monotop production have been carried out by the CMS and ATLAS experiments
at the LHC where no significant excess above the SM prediction has been observed,
and bounds are placed on the mass and couplings of the mediators \cite{Aad:2014wza, CMS:2016wdk, Sirunyan:2018gka}. 
It is notable that in addition to the FCNC interactions, the monotop production at the LHC is expected from
baryon number violation as well \cite{Dong:2011rh}.
While many of the above models have received much attention, 
considering the present experimental measurements tells us that these theories
are allowed providing that the new degrees of freedom are heavy enough 
or their couplings to the SM particles are significantly small.

In Ref.\cite{Hou:2017ozb}, the LHC signatures of the right-handed $tcZ'$ coupling have been studied. Such 
a FCNC coupling is inspired to explain the observed anomaly in the transitions of $B\rightarrow K^{(*)}$.

\begin{figure}[h!]
\centering 
\includegraphics[width=0.4\textwidth, height=0.25\textheight]{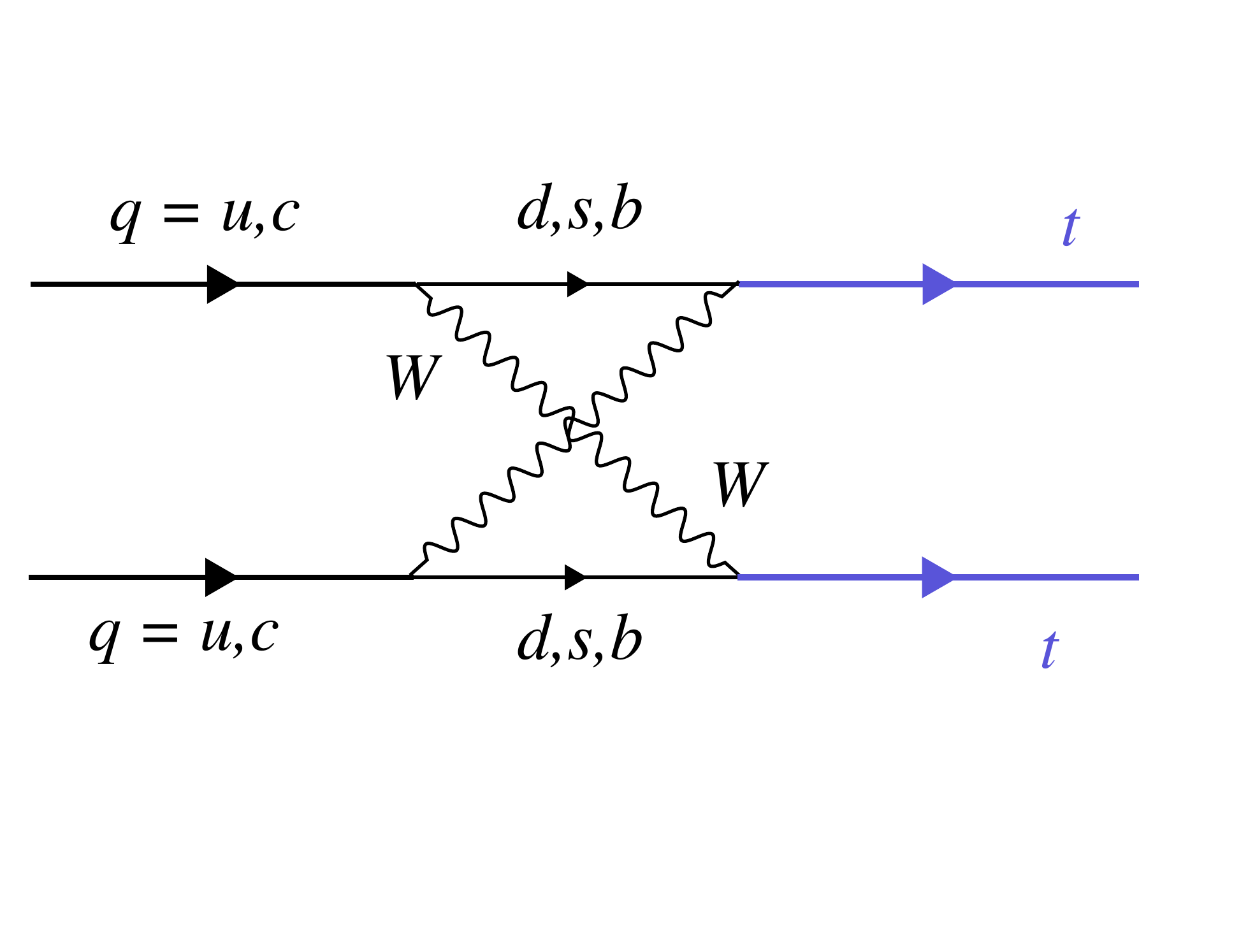}
\caption{Lowest order same-sign top quark pair production in the SM at the LHC. }
\label{fig:sm-tt}
\end{figure}

In addition to the monotop production, if a vector or a scalar flavor changing
exists, same-sign top quarks could be produced through
the $Z'$ or $\phi$ exchange in $qq\rightarrow tt$ ($\bar{q}\bar{q}\rightarrow \bar{t}\bar{t}$) process,
where $q = u,c$. 
In the SM, the same-sign top pair production proceeds through  one loop as depicted in Fig.\ref{fig:sm-tt}.
The SM $uu(\bar{u}\bar{u})\rightarrow tt (\bar{t}\bar{t})$ is proportional to 
$|V_{uq'}V_{tq'}|^{2}$ ($q' = d,s,b$) and the down-type quark masses. 
Therefore, within the SM the same-sign top pair production is expected to be significantly suppressed.
 This makes the same-sign top pair process a very interesting signature 
 as it has a very small amount of SM background and could be observed
as a pair of same-sign charged leptons associated with two $b$-jets.

Another clean signature for the top quark flavor changing of $tqZ'$ and $tq\phi$ at the LHC  is the same-sign top pair
production associated with a $W$ boson. 
In the SM framework, the lowest order $ttW^{-} (\bar{t}\bar{t}W^{+})$
production proceed through electroweak interactions at leading order with a production
rate of order of $10^{-4}$ fb.  Similar to the same-sign top pair, studying this process
has a much clear signature and this process is easy to discriminate from the SM background processes.
The  $ttW^{-}$ process could be observed 
in the form of two light flavor jet from the $W$ boson decay and a pair of same-sign charged leptons associated
with two $b$-jets from the semileptonic top quarks decays.

In this paper, we study  the same-sign top pair production at the LHC sensitivity on the basis of
a simplified model approach describing the top quark flavor changing through a scalar or a vector boson. 
New variables are introduced to 
enhance the signal-to-background ratio. 
We propose a transverse momentum based charge asymmetry which could indicate 
the new physics effects in $tt$ production process as well as angular observables
to distinguish between the signal from background processes and $tuX$ signal scenario 
from the $tcX$.
Additionally, we propose to use the $ttW^{-} + \bar{t}\bar{t}W^{+}$
process to search for the vector or scalar flavor violating effects at the LHC
as a complementary channel to the monotop and the same-sign top quark pair.
Then, we perform a statistical combination of $tt$ and $ttW$  processes
which improves the exclusion limits considerably. 
The analyses are based on a realistic simulation of the detector response and
the main SM background processes are taken into account. 
In particular, to include the detector response we focus on a CMS-like detector \cite{Chatrchyan:2008aa}.

The rest of this manuscript is organized as follows. 
In Section \ref{sec:model}, we present the theoretical formalism 
which is followed to search for the new physics effects. 
We then  extract the exclusion limits on the parameter space for both scalar and vector FCNCs
from the $D^0- \bar {D^0}$ mixing in Section \ref{d0d0bar}.
Section \ref{collider} is dedicated to present the details of the strategies and the analyses 
of $tt$ and $ttW$ channels as well as the new sensitive observables. The results of the
statistical combination of two processes are given in this section.
Our conclusions are presented in Section \ref{sec:conclusions}.

\section{Theoretical formalism}
\label{sec:model}

As we discussed in the previous section, the exchange of new top quark flavor changing states 
leads to tree level $tt$ and $ttW$ productions,
and consequently would result in modifications of the inclusive $tt$ and $ttW$ production cross sections of these processes.
We perform a detailed study of this possibility based on scenarios in which
the SM is extended by either adding a new vector boson or a new scalar boson affecting the 
$tt$ and $ttW$ productions at tree level. 
Similar to the SM, we assume QCD interactions respect 
flavor conservation and the FCNC
interactions occur via a weak sector.
The relevant effective Lagrangian including a new flavor changing vector ($Z'$) or a flavor changing scalar ($\phi$) 
is given by~\cite{Andrea:2011ws}: 
\beq 
\mathcal{L} = \mathcal{L}_{\rm{SM}}  +  \mathcal{L}_{\rm Kinetic} +   Z'_\mu  \bar u_i (a^{Z'}_{ij} \gamma^\mu + b^{Z'}_{ij} \gamma^\mu \gamma^{5})u_j 
+ \phi \bar u_i (a^\phi_{ij}  + b^\phi_{ij} \gamma^{5} ) u_j + \rm h.c. ,
\label{eq:lagrangian} 
\eeq
where the heavy mediators fields are denoted by $Z'$ and $\phi$, $\mathcal{L}_{\rm Kinetic}$ contains the kinetic 
terms of $Z'$ and $\phi$ fields.
The coupling matrices $a^{Z'/\phi}_{ij}$ and $b^{Z'/\phi}_{ij}$
denote the vector- and axial-couplings between up-type quarks with flavor $i$ and $j$ 
which proceeds through the exchange of either vector $Z'$ or scalar $\phi$.
In this work, for simplicity the axial couplings are neglected, {\it i.e.}  $b_{ij}^{Z'/\phi} = 0$, and we only consider the purely flavor changing 
interactions of the new states, {\it i.e.} $a_{ii} = 0$ for $i=u,c,t$.
The new states  $Z'$ and $\phi$ could play connection roles with an invisible fermionic state $\chi$ \cite{Kamenik:2011nb}.
In this paper, we restrict ourselves to only production mechanisms of $tt$ and $ttW$ involving $Z'$ and $\phi$. Consequently,
the interactions with the invisible sector are not taken into account. As we set $a_{ii} = 0$, any diagrams containing 
$Z't\bar{t}$ and $\phi t\bar{t}$ couplings do not contribute in the signal processes.

The phenomenology of  top quark flavor changing mediators has been discussed in many papers.
Some studies have considered the associate production of top quark with the mediator,
where the mediator subsequently either decays to another pair of SM particles or is a connector to 
Dark Matter (DM) and goes to a pair of DM particles ~\cite{Gupta:2010wt,Yanagida:1979gs}.
The production of a top quark in association with a charm quark at the LHC and the CLIC electron-positron collider 
via a $Z'$ flavor changing in the s-channel have been studied in Refs.~\cite{Arhrib:2006sg, Cakir:2010rs} and the discovery regions
of the new model parameter space have been presented. 
In addition to the production mechanisms to probe the properties of the new flavor changing mediators, 
such model's parameter space could be also probed by looking at the top quark decays.
Several papers have looked at $ t \to cX$ flavor changing transitions at the LHC
\cite{ Han:1995pk,  Yang:2008sb, Coimbra:2008qp, Aranda:2009rj,Khatibi:2015aal,Khatibi:2014via,Khanpour:2014xla, Malekhosseini:2018fgp}, 
mostly with the assumption of anomalous top-up(charm)-$g/\gamma/Z$ couplings. 
Nonetheless, in the case of assuming a new BSM particle, such a mediator is required to be heavier than the top quark~\cite{Langacker:2008yv}. 
However, as the bounds derived from these searches rely on flavor diagonal couplings, they do not apply to our particular scenario. 

\section{Indirect probes: $D^0- \bar {D^0}$ mixing}
\label{d0d0bar}

\begin{figure}[h!]
\centering 
\includegraphics[width=0.8\textwidth, height=0.2\textheight]{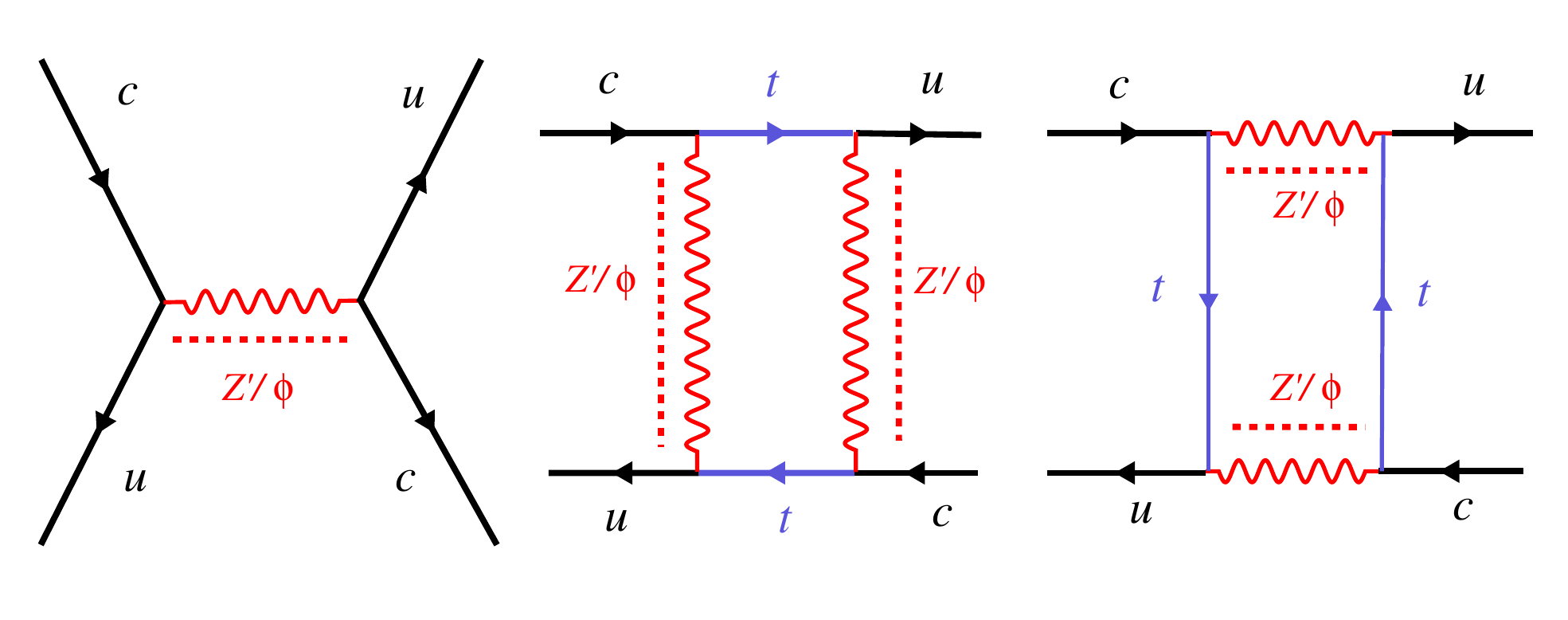}
\caption{The diagrams of $Z'/\phi$ that contributes to $D^0 -\bar{D^0}$ mixing at tree level and at loop level.  }
\label{fig:dmixing}
\end{figure}

In this section, we discuss the effects of a scalar $\phi$ and a vector $Z'$
boson with FCNC couplings to the up-type quarks on the
$D^0- \bar {D^0}$ mixing and derive the bounds on the couplings and masses
of scalar $\phi$ and $Z'$ boson.
Within the SM, the $D^0- \bar {D^0}$ mixing is a manifestation of FCNCs
that  occurs  as  the  flavor  eigenstates  are different  from  the  physical  mass  eigenstates  of
the $D^0- \bar {D^0}$ system.  Both the short-range quark-level transitions and the long-range
processes contribute to $D^0- \bar {D^0}$ oscillation.  The short-range contributions proceed via loops where 
the virtual particles are mediated \cite{Bianco:2003vb  , Burdman:2003rs, Artuso:2008vf}.  
This causes the study of the $D^0- \bar {D^0}$ mixing more interesting as 
new physics models with new (FCNC) degrees of freedom could be examined through it.
The parameters of the new vector and scalar FCNC interactions to two up-type quarks could be
significantly constrained using the measurement of $D^0- \bar {D^0}$ mixing
which is affected at both the tree level and loop level.
The Feynman diagrams of a scalar $\phi$ or a vector $Z'$ flavor changing which contribute to the $D^0- \bar {D^0}$ mixing
are depicted in Fig.\ref{fig:dmixing}.

 Assuming the mediator has up-charm $(a_{uc})$, up-top $(a_{ut})$, 
and charm-top $(a_{ct})$ couplings, it can have significant contribution to neutral $D$ meson mixing. 
 If the mediator is $Z'$, its contribution 
 to the mass difference between the two mass eigenstates is given by \cite{Arhrib:2006sg}:  
\begin{align}
\label{dmixing}
\Delta M_D &= \frac{ f_D^2 M_D^2 B_D}{12 m_{Z'}^2}\left[a_{uc}^2+ (a_{ut}a_{ct})^2 \frac{x}{8 \pi^2}(32 f_{_{Z'}}(x)- 5 g_{_{Z'}}(x))\right]\\
\text{where } &f_{_{Z'}}(x)= \frac{1}{2}\frac{1}{(1-x)^3}[1-x^2+2x \log x]\nonumber\\
&g_{_{Z'}}(x)= \frac{2}{(1-x)^3}[2(1-x)+(1+x)\log x]\nonumber
\end{align}
and the scalar FCNC contributions to the mass difference can be expressed as \cite{Golowich:2007ka}:
\begin{align}\label{dmixing1}
\Delta M_D &= \frac{ f_D^2 M_D B_D}{24 \pi^2 m_{\phi}^2}\left[a_{uc}^2+ (a_{ut}a_{ct})^2 f_\phi\left(\frac{m_t^2}{m_\phi^2}\right)\right]\\
\text{where } &f_\phi(x)= - \frac{1}{1-x} - \frac{\log (x)}{(1-x)^2} + \frac{ x^2 -4x + 3 +2 \log x}{2(1-x)^3} \nonumber. 
\end{align}
where in both Eq.\ref{dmixing} and Eq.\ref{dmixing1}, the first term in the square bracket represents the tree level 
contribution and the second term is due to the box contribution, 
$x\equiv m_{Z'/\phi}^2/m_t^2$, $M_D\sim 1.9\ \gev$ is the mass of $D$ meson, 
$f_D\sim  223\ \mev$ is its decay constant, $B_D\sim 1$ is the bag model parameter, 
and $a_{ij}$ is the coupling between $i$ and $j$ up-type quarks. In this work, we set 
$a_{uc} = 0$ and only focus on the FCNCs of top and up/charm quarks via $Z'$ and $\phi$. 
The current bound on $D$ meson mixing is $ \Delta M_D < 3.6 \times 10^{-7}$~\cite{delAmoSanchez:2010xz, Staric:2007dt}
by which constraints in the planes of $(a_{ut},m_{Z'/\phi})$  and 
 $(a_{ct},m_{Z'/\phi})$ are presented in Fig.\ref{fig:bounds}  for different values of $a_{ct}$ and $a_{ut}$, respectively.
 The bounds from $D$ meson mixing on  $a_{ut}$ and $a_{ct}$  are similar.

\section{Collider signatures}
\label{collider}

At the LHC, the rates of
top quark production are of the order of several hundreds pb, which are relatively
large among the interesting processes.
Although top quarks are mainly produced singly via electroweak interaction or 
in pair of top-antitop via strong interactions, searches for 
same-sign top quarks and same-sign top quarks associated with a $W$ boson are important as
new physics can enhance their rates. 

In this section, first we will carry the analysis in the same-sign top quark pair production, and find the allowed region of the
parameter space for different scenarios of the integrated luminosities at the LHC with the center-of-mass
energy of 14 TeV. Then, the reach with same-sign top quarks plus a $W$ boson is studied. Finally, a combined statistical analysis is performed on 
$tt$ and $ttW$ channels and we show how the combination can extend the sensitivity in the parameter space.

\subsection{Two same-sign tops }
\label{sec:2sst}

\begin{figure}[h!]
\centering 
\includegraphics[width=0.5\textwidth, height=0.12\textheight]{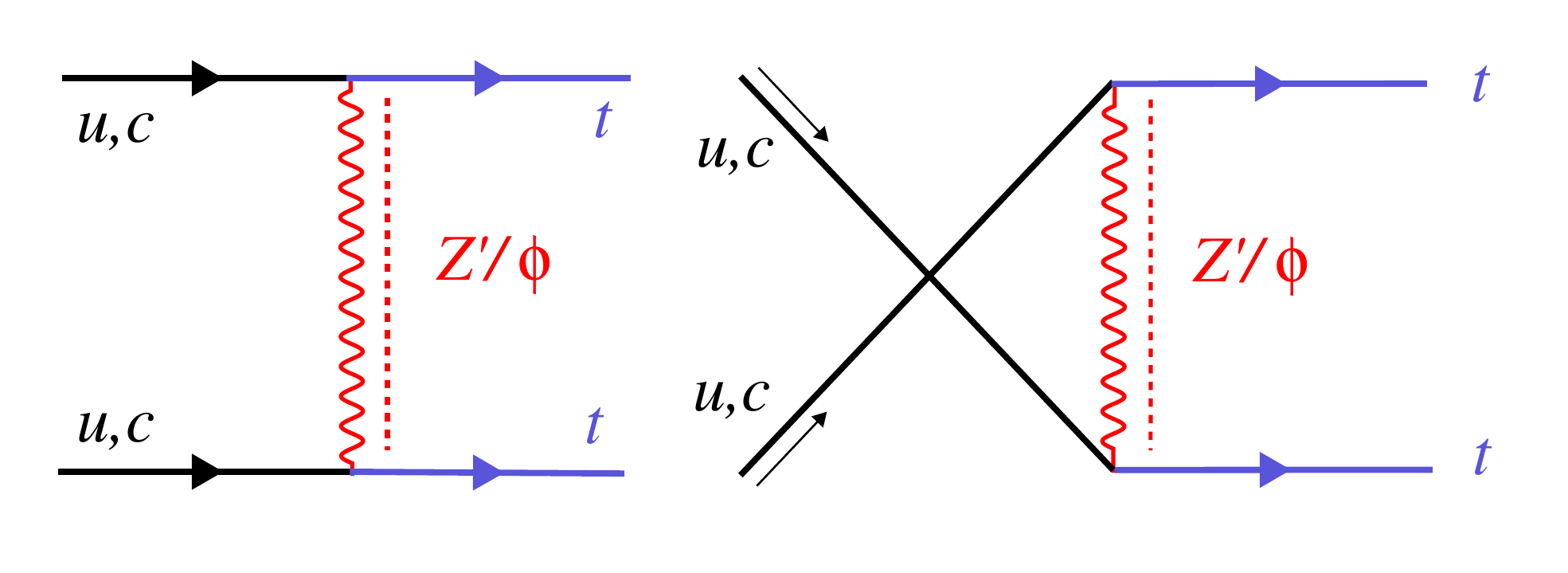}
\caption{Representative Feynman diagrams for the same-sign top quark pair production at the LHC  via  a  scalar ($\phi$) and a vector ($Z'$). }
\label{fig:tt-feynman}
\end{figure}

The existence of flavor changing in the top quark sector via 
either a scalar $\phi$ or a vector $Z'$ would allow the production 
of same-sign top quark pair in proton-proton collisions at the LHC. 
Representative Feynman diagrams at parton level for the same-sign top quark 
pair production at the LHC  via  a flavor changing scalar ($\phi$) or a vector boson ($Z'$) exchange  are
shown in Fig.\ref{fig:tt-feynman}.

\begin{figure}[h!]
\centering 
\includegraphics[width=0.7\textwidth, height=0.2\textheight]{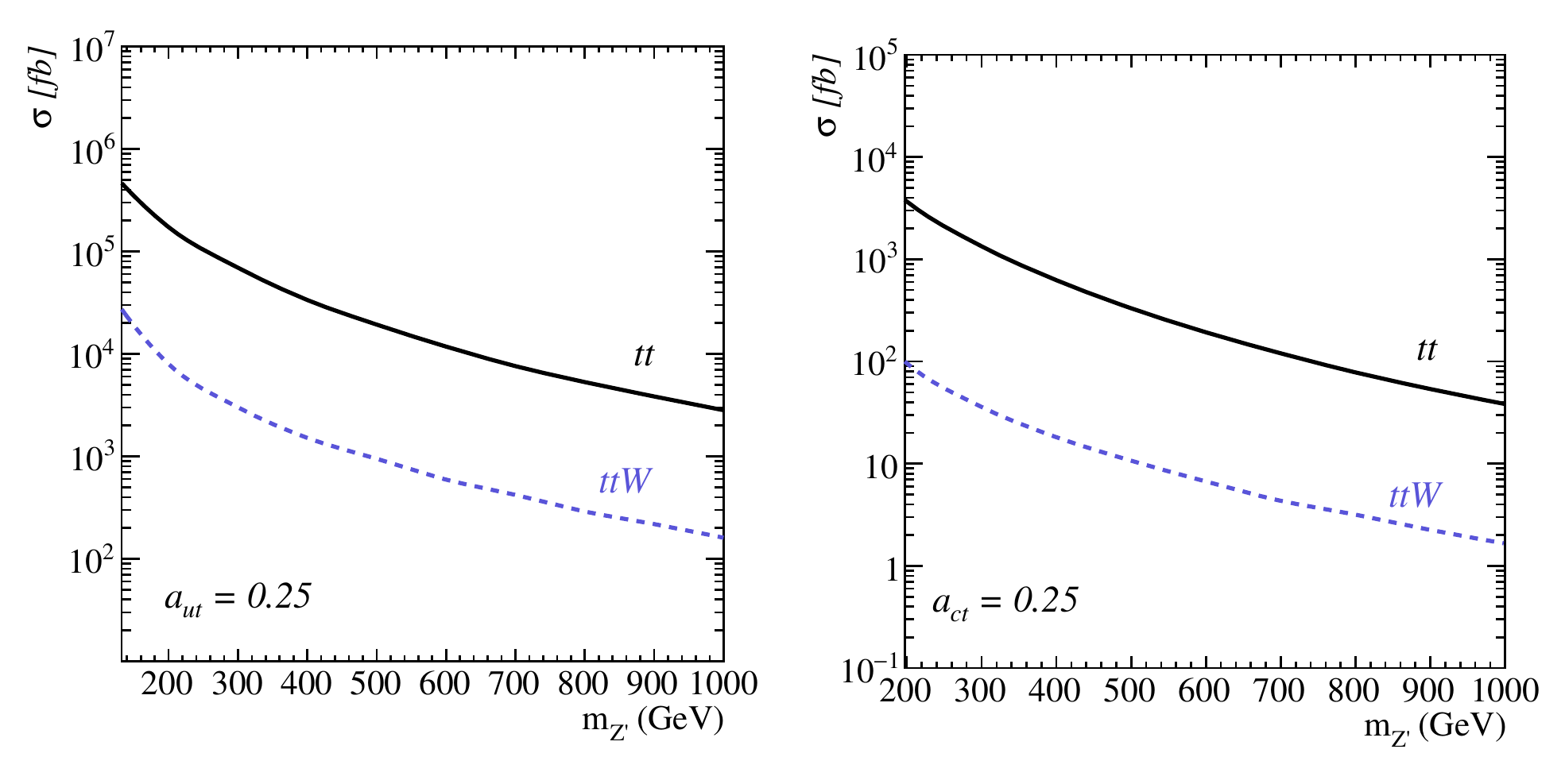}
\includegraphics[width=0.7\textwidth, height=0.2\textheight]{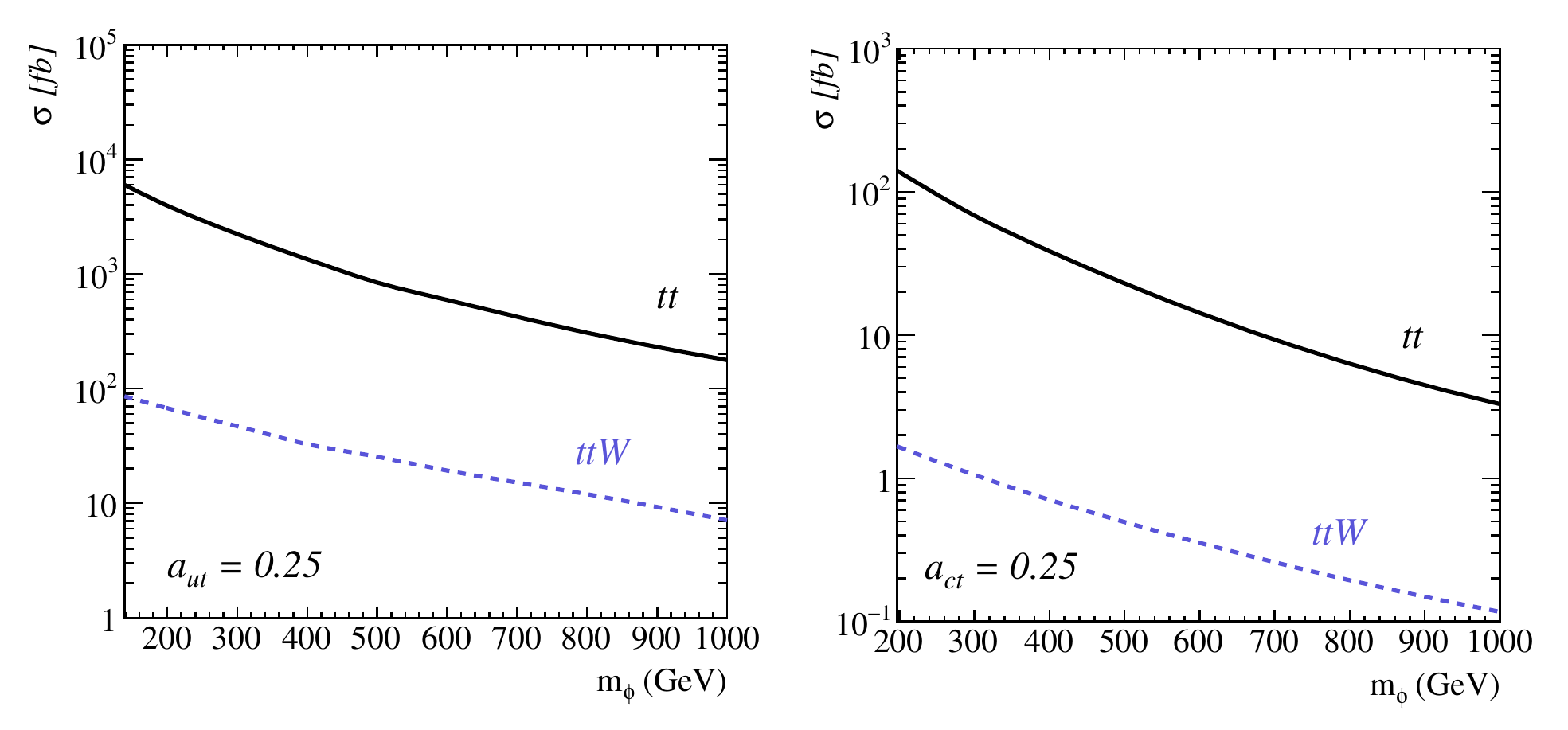}
\caption{The production cross sections of the same-sign top pair and same-sign top pair associated with a 
$W$ boson in terms of the mass of the $Z'$ (top) and $\phi$ (bottom). The right plots are for the case $a_{ut} = 0.25, a_{ct} = 0.0$ and the left plots 
are for the case that $a_{ut} = 0.0, a_{ct} = 0.25$.}
\label{xsec}
\end{figure}

Same-sign top pair production is the most optimal channel to look for our signal, because of two main reasons:
({\it i}) it has a very low and reducible background; ({\it ii}) in the case of up-top flavor changing, the top quarks will be 
produced from up quarks, which are proton's valence quarks. Due to their higher PDFs, 
the cross section involving up quarks is more significant, and the top quarks in the final state are more energetic. 
This signal property is helpful for more discrimination of signal from background.
Top anti-quarks coming from $\bar{u}$-quarks (sea quarks) suffer from lower PDF, and the $\sqrt{\hat s}$ 
in this case tends to be much lower. Therefore, the final states of anti-tops are less energetic with respect to the same-sign top pair. 
The production cross sections of the $tt$ process as a function of the $Z'$ and $\phi$ mass once with  $a_{ut} = 0.25$ 
and once with $a_{ct} = 0.25$ are presented in Fig.\ref{xsec}.

As indicated before, study of the  same-sign top quark due to the scalar and vector FCNCs has the advantage of enhancing the sensitivity 
to the signal parameter space. Since the jets charge measurement at the LHC is a much complicated task and has 
high uncertainties, we require top quarks to decay leptonically, even though lower statistics is expected due to lower 
branching fractions. The final state of signal consists of two same-sign charged leptons, two $b$-jets and
large missing energy. Based on the final state, the main backgrounds are listed below which some emerge as a result of detector limitations
such as missing some objects at the detector, or fake objects:
\begin{eqnarray}
&&pp \rightarrow   tt (\bar{t}\bar{t}) \rightarrow   \ell^\pm \ell^\pm b b \nu_{\ell} \bar{\nu}_{\ell}~\text{(SM process)}, \nonumber\\
&&pp \rightarrow   W^\pm W^\pm |_{\text{leptonic decay}} + \text{jets}, \nonumber\\
&&pp \rightarrow    t \bar{t} W^\pm |_{\text{leptonic decay}},  \text{and}~ pp \rightarrow    t \bar{t} Z |_{\text{leptonic decay}}, \nonumber \\
&&pp\rightarrow    W^\pm W^\pm  |_{\text{leptonic decay}} + \text{jets~,~Double Parton Scattering}, \nonumber\\
&&pp\rightarrow     t \bar{t}  \rightarrow  b( \to \ell)\ell \nu_\ell\ \bar b j j, \nonumber \\
&&pp \rightarrow  W^{\pm}W^{\mp} jj , \text{and} ~ pp \rightarrow  V V jj , ~V = Z, \gamma,\nonumber
\label{eqn:backgrounds}
\end{eqnarray}

\begin{figure}[h!]
\centering 
\includegraphics[width=0.95\textwidth, height=0.2\textheight]{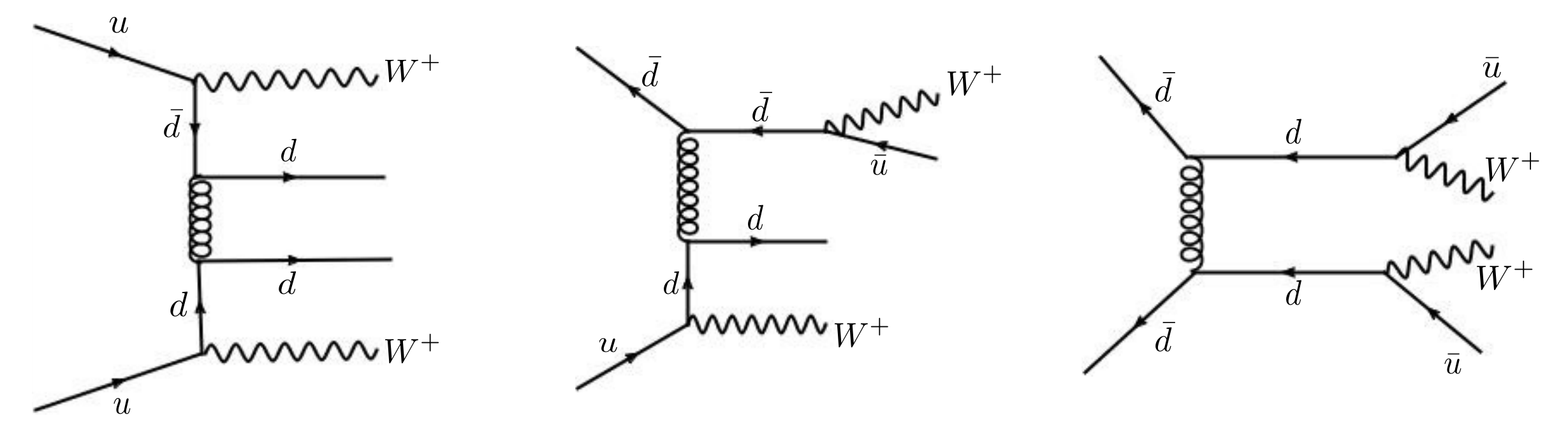}
\caption{Representative Feynman diagrams for the $W^{\pm}W^{\pm}$+jets production at the LHC.
 The initial states in this background can be both from the valence quarks (left),
  only one of them coming from the valence quark (middle), or both coming from sea quarks (right).}
\label{fig:bkg}
\end{figure}

The production of double same-sign top quarks $ tt (\bar{t}\bar{t})$ in the SM are through loops involving CKM entries and thus 
negligible. The $ p p \to W^{\pm}W^{\pm}$+jets process is the irreducible background, 
where the $W$s are same-sign and they both decay leptonically.  Figure \ref{fig:bkg} depicts
 example Feynman diagrams for the QCD induced production of $W^{\pm}W^{\pm}$+jets. 
 
 The $t\bar{t}W^{\pm}$ \cite{Campbell:2012dh} and $t\bar{t}Z$ processes, 
 with leptonic decay of the $W$ boson or $Z$ boson and semi-leptonic 
 decay of the $t\bar{t}$ pair, could have same-sign dilepton in the final state therefore it contributes to the background. The final state of these
 two processes contains more jets with respect to the $tt(\bar{t}\bar{t})$ signal events.
 The  $t\bar{t}Z$ process is in particular a background when one of the charged leptons from the $Z$ boson 
 decay escapes detection.
 
Another important background is $ p p \to t \bar t \to  j_{b} \ell \nu_\ell\ \bar j_{b} j j $, 
where one of the $b$-jets ($j_{b}$) include a lepton in its jet (the $B$ meson decays to  leptonic final states with s branching fraction of about $10\%$). 
If the two leptons in the process have same-signs, this process would be a background to our signal. 
Although the odds of such process seem small, due to the high cross section of $ t \bar t$  at the LHC, this process could be  important. 

The other backgrounds are due to detector mis-measurements or objects that are missed in the detector. 
Some of these are $W^{\pm}W^{\mp}$+jets and $VV$+jets, where the $V= Z$ or $\gamma$ decay leptonically. 
In such situations, we may have two leptons of the same-sign and two jets. 
This happens in the cases that a real opposite-sign dilepton pair is present in the final state 
and the charge of the leptons is mistaken by the detector.
The charge misidentification probability is negligible for muons with respect to electrons.
 For the electron, the main effect comes from the conversion of  $e^{\pm} \rightarrow e^{\pm}\gamma \rightarrow e^{\pm}e^{\mp}e^{\pm}$
 in the detector.  The charge misidentification probability is dependent on the lepton $p_{T}$
 and goes up at very high $p_{T}$. For the processes considered here, 
highly boosted leptons are very unlikely. Hence, due to small cross sections of these processes
and low charge misidentification of the leptons, we ignore these backgrounds in the analysis. 

Another source of background is the $W^{\pm}W^{\pm}jj$ from \textit{Double Parton Scattering (DPS)}.
The $W^{\pm}W^{\pm}$+jets from DPS arises from the cases that two various hard parton-parton interactions
occur at the same time in a single proton-proton collision.
The rate of $W^{\pm}W^{\pm}jj$ is estimated using $\sigma(W^{\pm}j)^{2}/(2\sigma_{eff})$ \cite{Kulesza:1999zh}, where $\sigma(W^{\pm}j)$
is the $W^{\pm}$+jet rate, factor of two in the denominator is a symmetry factor for the identical processes 
and $\sigma_{eff}$  is the total effective proton-proton cross section at the LHC which is of the order of
15 mb. 

The signal  processes are simulated using the Monte Carlo (MC) generator
 {\tt MadGraph5-aMC@NLO} \cite{Alwall:2011uj},
 and an already available Universal FeynRules Output (UFO) 
 model~\cite{Alloul:2013bka,Degrande:2011ua}.
 Then the parton-level events are passed through {\tt Pythia 6}~\cite{Sjostrand:2007gs} for 
 parton shower, hadronization and decay of unstable particles. 
 The detector effects are simulated using {\tt Delphes 3}~\cite{deFavereau:2013fsa}. 
 At the detector level, the pile-up effects are not considered. The background 
 samples are also generated in a similar fashion. 
 For $ t \bar t$ background, in ${\tt Pythia}$, the mesons which include $b$-quarks are forced to decay leptonically to increase the statistics of our study. 
 
The considered final states are $\mu^{\pm}\mu^{\pm} \nu_{\mu}\nu_{\mu} j_{b}j_{b}$, $e^{\pm}e^{\pm} \nu_{e}\nu_{e} j_{b}j_{b}$,
and $e^{\pm}\mu^{\pm} \nu_{e}\nu_{\mu} j_{b}j_{b}$, where $j_{b}$ denotes a jet originating from the hadronization of
a $b$-quark. 
The event selection is designed to identify same-sign charged lepton events compatible with the two same-sign top quarks events, while
keeping down the contribution of background processes. To trigger the events, one can either rely on the 
single or double lepton triggers \cite{Khachatryan:2016kod}. These triggers are based on the presence of a single energetic isolated lepton or
two low $p_{T}$ isolated charged lepton.
To ensure the events satisfy the trigger, it is required to have two same-sign charged leptons with 
 $p_T(\ell) > 25$ and $ \eta < 2.5$. Furthermore, the leptons are required to satisfy an isolation criteria. The relative isolation
 for the electrons and muons are defined as:
\beq\label{iso}
\text{RelIso} (\ell) = \frac{ \sum_{i}^{\Delta R (\ell,i ) < 0.3 }  p_T(i)  }{p_T (\ell)}, 
\eeq 
where the sum in the numerator is over the transverse momenta of 
particles lying inside a cone with a radius of $R = 0.3$ except for the charged lepton $\ell$ itself.
For an isolated lepton, the $\text{RelIso} (\ell) $ is expected to take small values close to zero. In this analysis,
the maximum value of 0.15 is taken.
Jets are reconstructed with the anti-$k_{T}$ 
algorithm \cite{Cacciari:2008gp} as implemented in the {\tt FastJet} package \cite{Cacciari:2011ma} with a distance parameter of 0.4.
Each event is required to have at least two jets with $p_{T}(j) > 30 $ GeV and $|\eta_{j}| < 2.5$ from which at least one 
is required to be $b$-tagged. The efficiency of $b$-tagging  and the misidentification rates are 
dependent on the jet $p_{T}$ and are assumed to be similar to  the CMS detector \cite{CMS:2016wdk}. At a $p_{T}$ of 40 GeV,
the $b$-tagging efficiency is  $60\%$, $c$-jet misidentification rate is $14.6\%$, and a misidentification rate for light flavor jet is $1.1\%$ \cite{Chatrchyan:2012jua}.
In order to have a well isolated objects in the final state, the angular separation between all selected objects to be larger than $0.4$, at the detector level. 
All the explained cuts above except for the isolation cut on the charged leptons and $b$-tagging are called as the basic cuts.
The efficiencies after each cut for three signal scenarios and for the main backgrounds are shown in Table~\ref{tab:eff}.
As it is shown in Table~\ref{tab:eff}, the isolation requirement does not affect the signal or the $W^{\pm}W^{\pm}$+jets, $t\bar{t}W^{\pm}$, and $t\bar{t}Z$ backgrounds, 
but it completely eradicates the $t \bar t$ background. 
That is because in the $t \bar t$ background, one of the same-sign leptons comes from a $b$-jet and thus the environment 
around the lepton is polluted by other particles in the jet, whereas in other processes the leptons are produced 
isolated and they stay isolated throughout the detector.  
The requirement of having at least one $b$-tagged jet is quite effective to suppress the $W^{\pm}W^{\pm}$+jets background.
Technically, the jets in $W^{\pm}W^{\pm}$+jets could be $b$-jets, but those are produced either from 
 $b$-quark PDFs or from CKM flavor changing couplings which both are suppressed. 
As can be seen in Table~\ref{tab:eff}, this requirement significantly reduces 
the $W^{\pm}W^{\pm}$+jets background while leaving the signal almost unchanged.

\begin{table}[h!]
\begin{center}
\begin{tabular}{l c| c c c |c c c c c}
\hline
\hline
cuts & & \multicolumn{3}{c}{$m_{Z'}$} & &  \multicolumn{2}{c}{SM bkg}\\
\hline
 &    & ~~$600\ \gev$ & ~~$1\ \tev$&& ~~~$W^{\pm}W^{\pm}$+jets& ~~$ t\bar t$ & ~~$ t\bar t W^{\pm}$  & ~~$ t\bar t Z$   \\
\hline
Basic cuts  &&  ~~$14.0\%$&~~~ $13.1\%$  &  &~~~~ $15.0\%$& ~~~$20\%$ &  $31\%$  &  $8\%$  \\
Isolated leptons &&~~ $ 13.1\%$  & ~~~$12.3\%$  &  &~~~~ $ 14.5\%$ &~~~ $0.0\%$  &  $29\%$  &  $7\%$ \\
B-tagging &&~~  $10.6\%$&~~~ $10.0\%$&   & ~~~~$1.2\%$ & ~~~-- &  $20\%$ &  $4\%$   \\
$\Delta \phi (\ell_1, \ell_2 ) > 1.5 $ &&~~ $9.6\%$& ~~~$9.2\%$ & & ~~~~$0.7 \%$&~~~--  &  $11\%$ &  $3\%$ \\
\hline
\hline
\end{tabular}
\caption{The efficiencies of each cut on two benchmark points $m_{Z'} = 600 \ \gev$ and $1\ \tev$ and the main backgrounds. 
The couplings of the benchmarks are not important because there is no interference between the signal and background;
 thus the behavior of the distributions do not depend on the couplings. As shown in the table, the background 
 $t \bar t$ gets suppressed by the lepton isolation cut, and the $W^{\pm}W^{\pm}$+jets is reduced significantly by the $b$-tagging
requirement as well as the $\Delta \phi (\ell_1, \ell_2)$ cut. }
\label{tab:eff}
\end{center}
\end{table}

The cross section of the $W^{\pm}W^{\pm}jj$ from DPS after the basic cuts is found to be quite negligible as a result it is not considered
in the rest of this analysis.

\begin{figure}[h!]
\centering 
\includegraphics[width=0.5\textwidth, height=0.25\textheight]{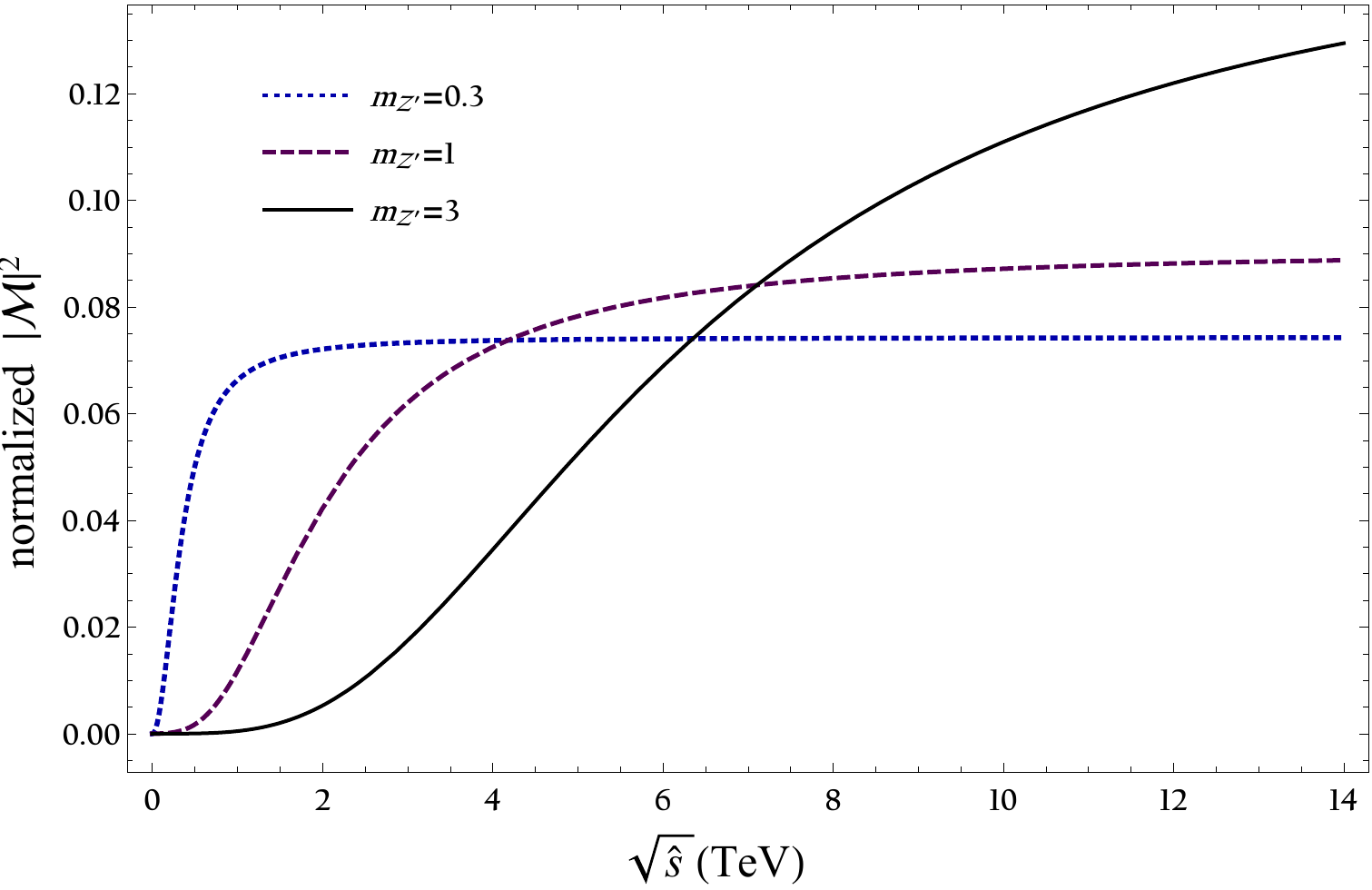}
\caption{The behavior of squared matrix element in terms of $\sqrt{\hat{s}}$ for the signal process $uu\rightarrow t t$ for $ m_{Z'} = 0.3, 1, 3$ 
TeV assuming the scattering angle in the center-of-mass frame equals to $\pi/2$.}
\label{me}
\end{figure}

To further enhance the signal and suppress the main background events, one can look for the special features in the kinematics of the signal.
In Appendix \ref{app:xsection}, we show the analytical expression of the $\left| \mathcal{M}\right|^2_{u u \to t t}$.  
Because none of the backgrounds have exactly the same initial and final states as the $tt$ signal, there is no interference between
 signal and background at tree level. Therefore, the squared matrix element scales as $a_{ij}^4$ and none of the kinematics should 
 depend on the coupling.  Consequently, when the benchmarks are specified, only the mass of $Z'$ is important.  
 In Fig.\ref{me}, the normalized squared matrix element 
 of $uu\rightarrow tt$ versus $\sqrt{\hat{s}}$ for three masses of $Z'$ are shown for instance for the case of $\cos\theta = 0$
 , where $\theta$ is the scattering angle in the partonic center-of-mass frame.
 As one can see, for larger $m_{Z'}$, the $\sqrt{\hat s}$ becomes more important. This is a common feature in the $s$-channel processes, 
 but not so trivial in $t$- and $u$-channels. The $\hat s$ for the $tt$ production
 could take large values because the initial states are valence quarks. Since the squared matrix element dictates 
  the behavior of the final state distributions, the large $ \hat s$ seems to be more important in 
  processes with large $m_{Z'}$. As a result, for larger $m_{Z'}$, the final state top quarks are more energetic, 
  and their final states are more collinear.  Since  both top quarks are back-to-back, their final 
  states are expected to be almost back-to-back as well. 
  
\begin{figure}[h!]
\centering 
\includegraphics[width=0.5\textwidth, height=0.25\textheight]{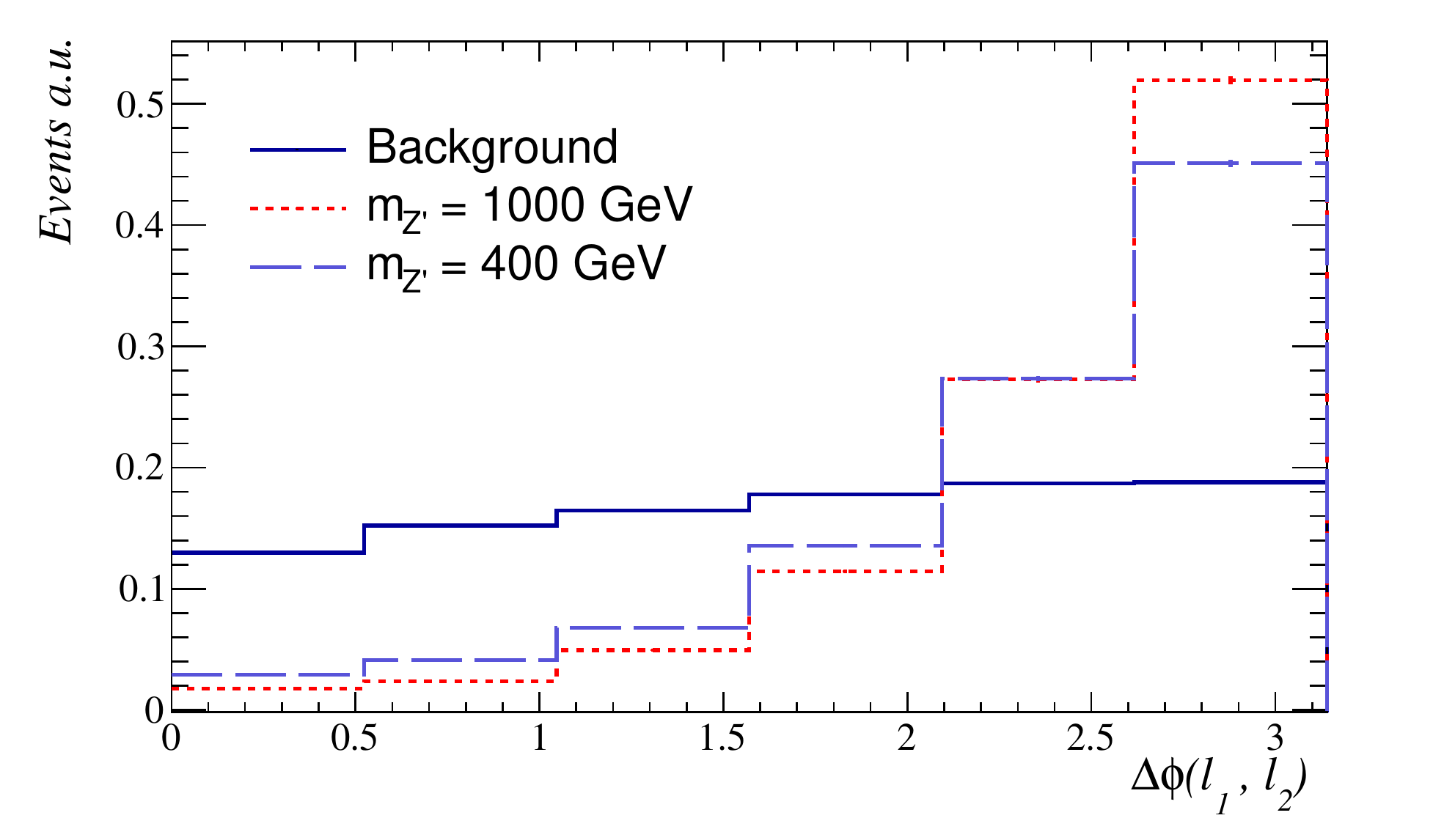}
\caption{The distribution of $\Delta \phi (\ell_1, \ell_2)$ for two benchmark points and sum of all background processes.
 The leptons that come from heavier mediator tend to be more back-to-back.}
\label{fig:deltaphi}
\end{figure}

  Figure \ref{fig:deltaphi} shows the $\Delta \phi (\ell_{1}, \ell_{2})$ and indicates that 
   the final decay products of top quarks tend to become more back to back as $m_{Z'}$ increases. 
  This is while the background shows no interesting feature for all values of $\Delta \phi$. 
  Thereby, with a cut on $ \Delta \phi$ we can enhance our signal over background discrimination.  
  Another variable that gets affected by this observation is $H_T = \sum_{\text{visible}} p_T$. 
  For larger values of  $m_{Z'}$, the distribution of $H_T$ peaks at higher value. 
  However, a cut on $H_T$ is not very efficient and it is $m_{Z'}$ dependent as a consequence no cut is applied here. 

\begin{figure}[h!]
\centering 
\includegraphics[scale=0.35]{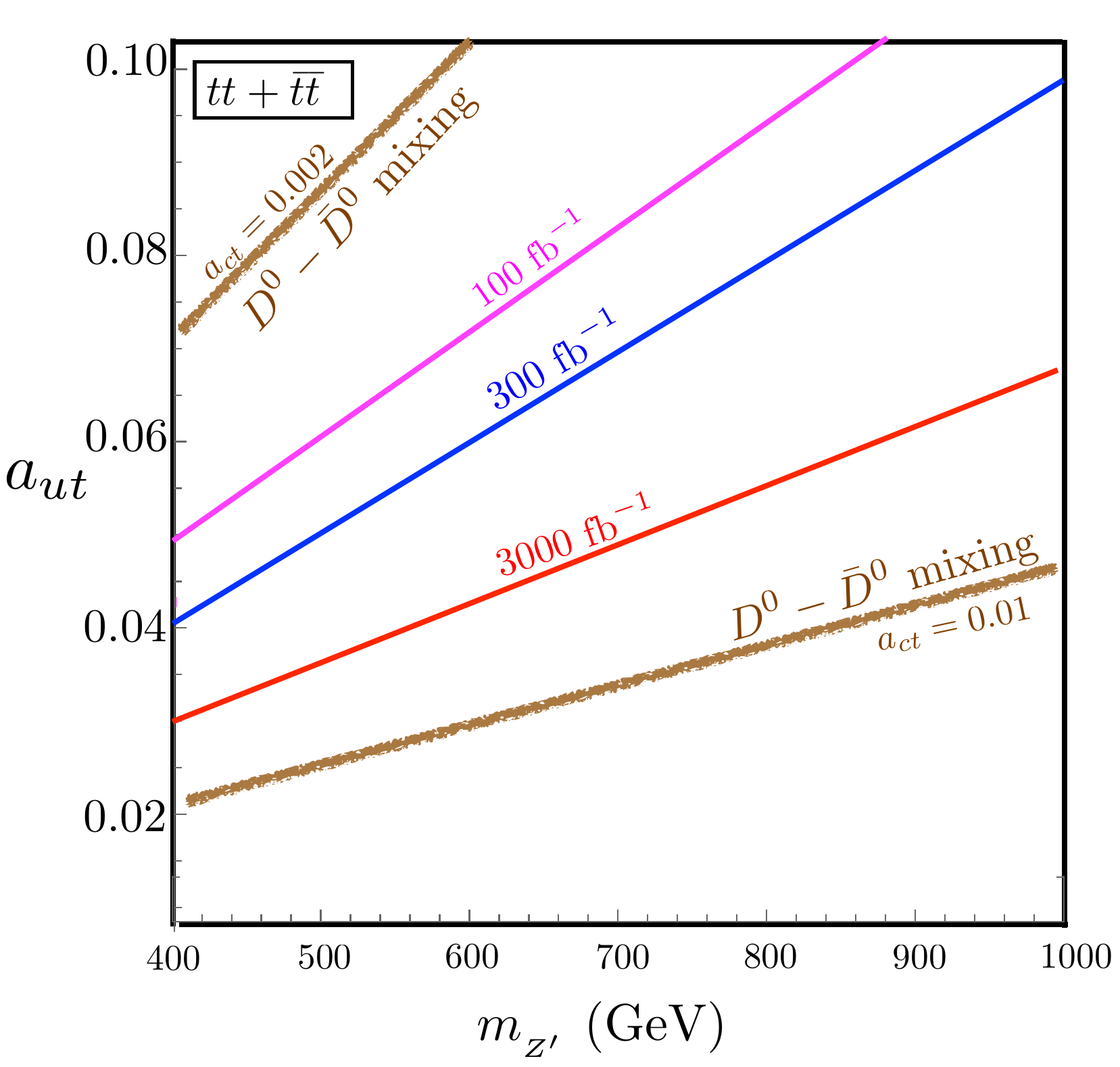}
\includegraphics[scale=0.35]{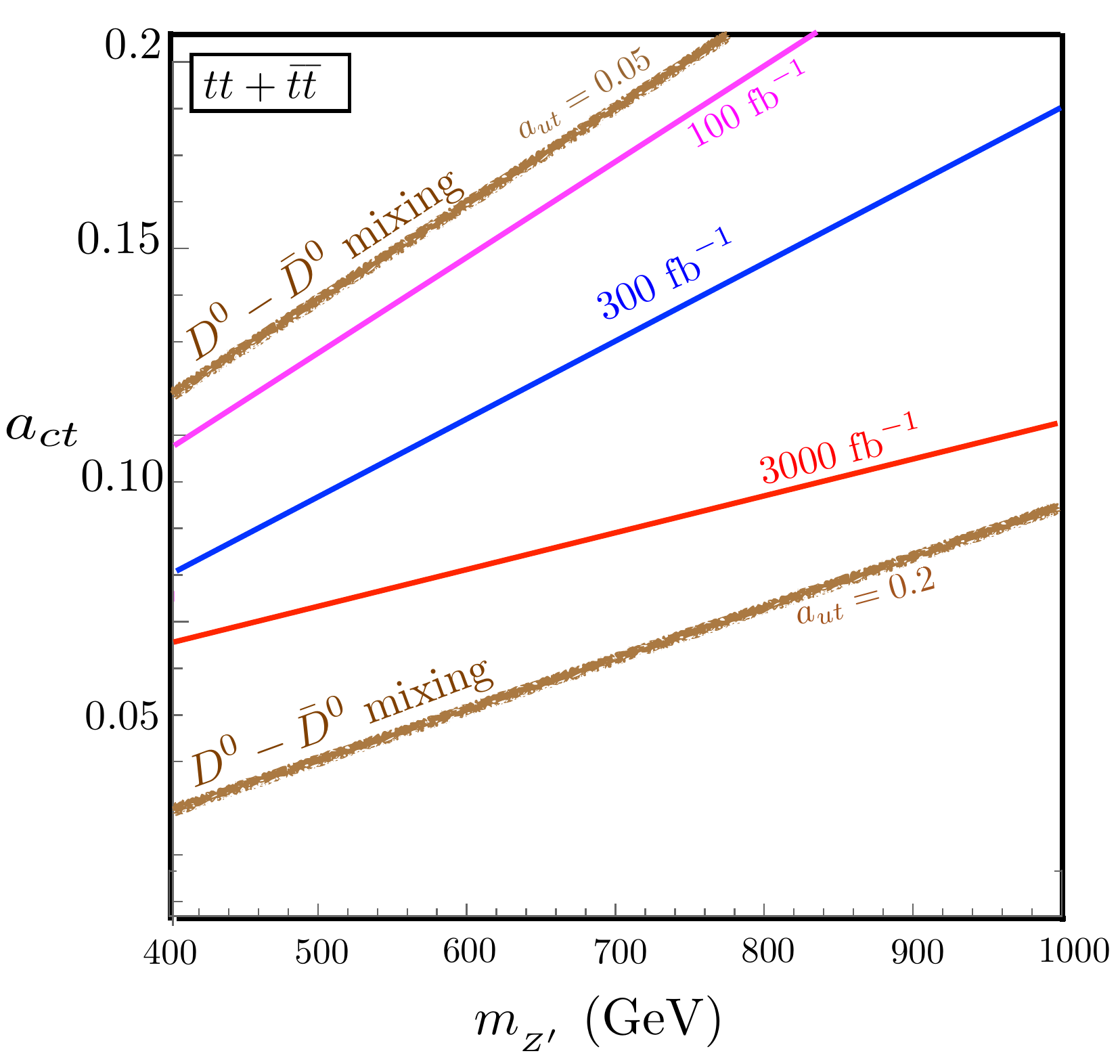}\\
\includegraphics[scale=0.35]{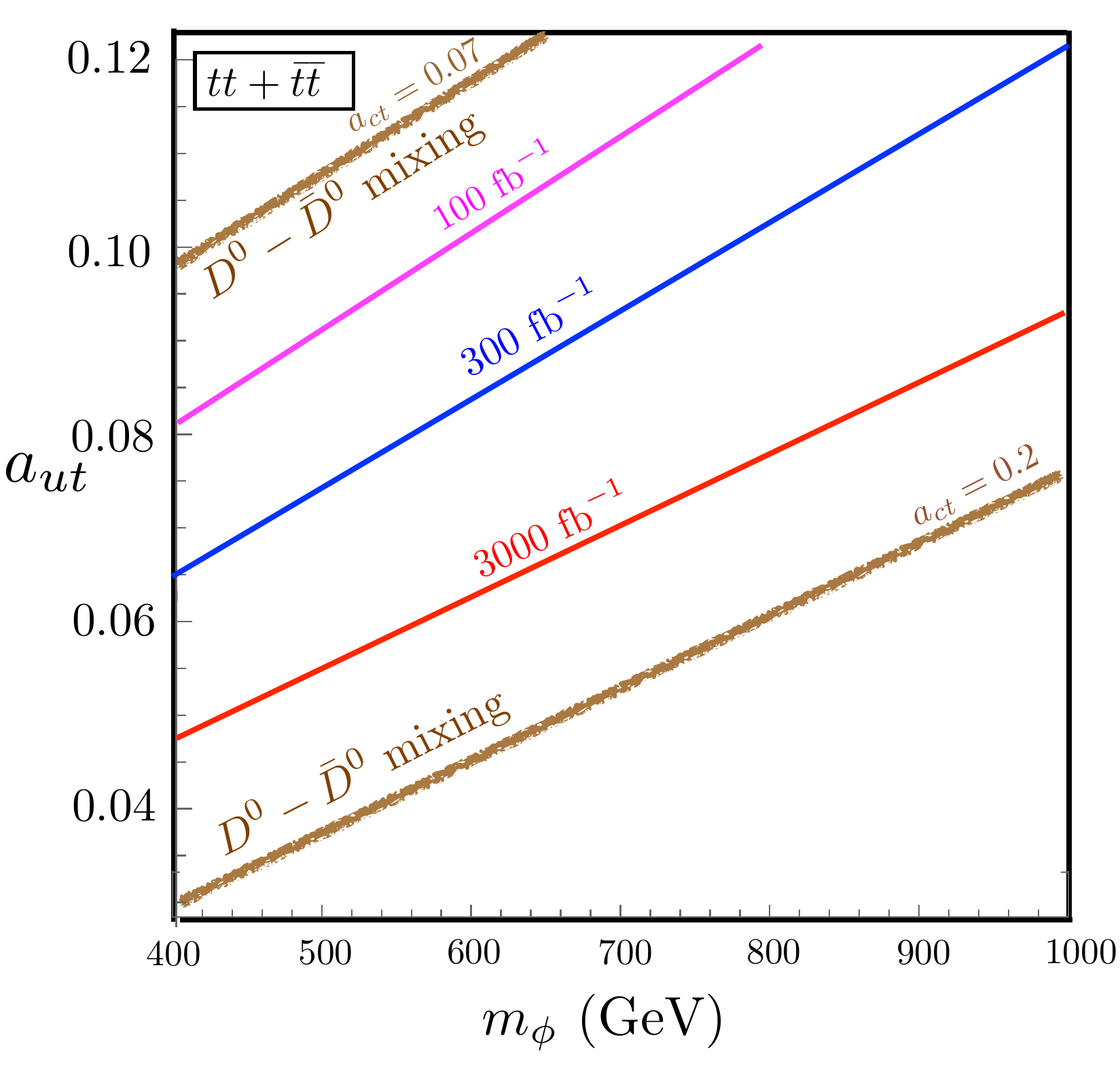}
\includegraphics[scale=0.35]{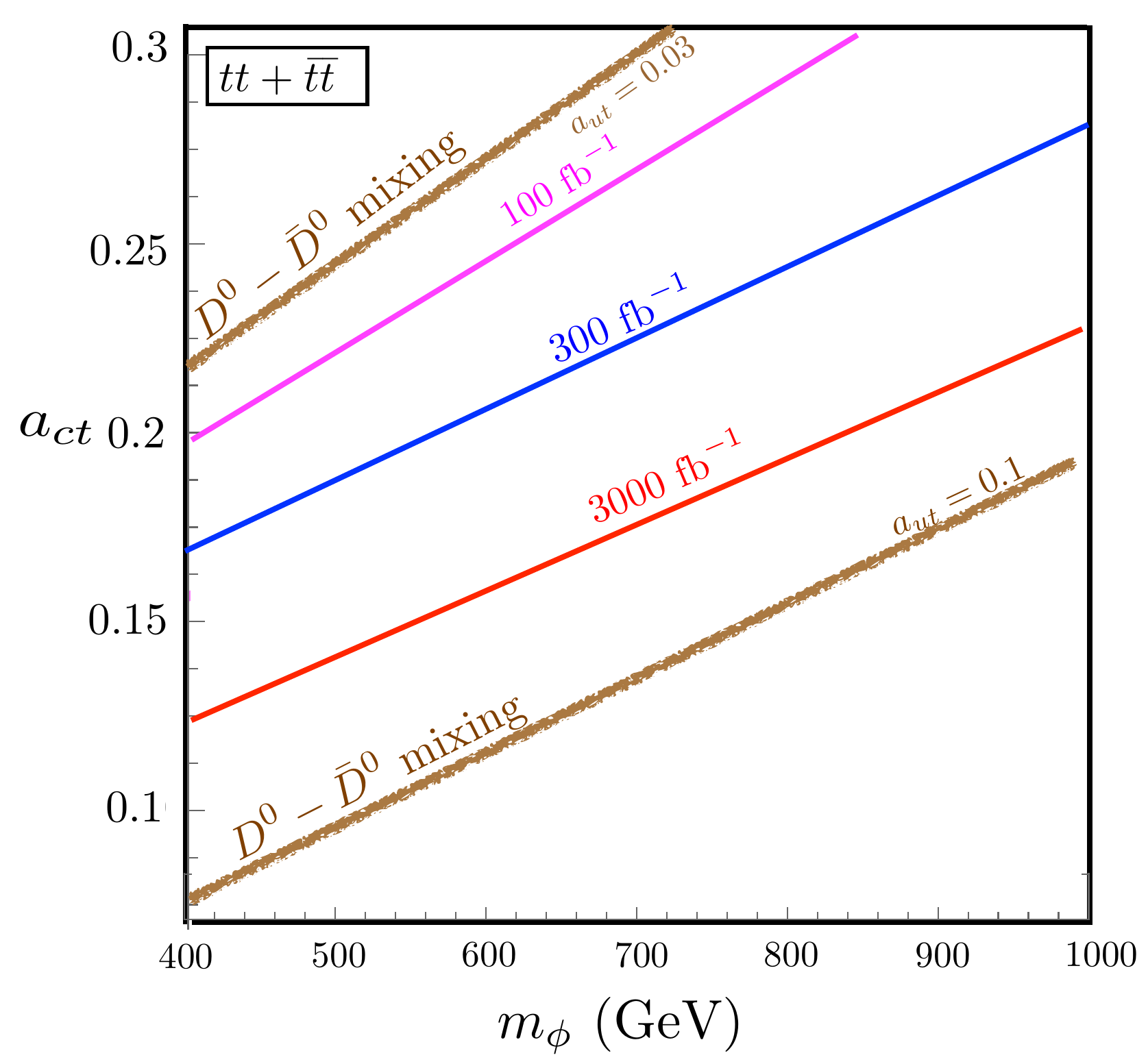}\\
\caption{The $95\%$ CL limits on $Z'$ (up) and $\phi$ (down) that have flavor changing coupling with up and top quarks (left) and with charm and top quarks (right). 
The region to the left of the plots are excluded up to $95\%$ confidence level with $ 100 \ \fb^{-1}$ (magenta), $300 \ \fb^{-1}$ (blue) 
and $3000 \ \fb^{-1}$ (red) integrated luminosity. The brown lines represents the $ D^0 - \bar D^0$ mixing, where $ a_{uc} = 0$, 
but  to get any bounds from neutral D mixing, we need to have both $ a_{ut} \neq 0$ and  $a_{ct} \neq 0$. 
 For the LHC bounds, we have assumed only one flavor changing couplings: $a_{ct} = 0$ (left) and $a_{ut} = 0$ (right),
  but for the neutral D mixing, we have to turn both couplings on.   }
\label{fig:bounds}
\end{figure}

To quantify our sensitivity,  a single
bin counting experiment is used to set
the limits.  A Poisson distribution is taken as the probability of measuring
$n$ events:
\begin{eqnarray}\label{poisson}
\mathcal{P}(n | n_{s}, n_{b})  = e^{-(n_{b}+n_{s})} \times \frac{(n_{b}+n_{s})^{n}}{n!},
\end{eqnarray}
where $n_{s} = \epsilon\times \mathcal{L}\times\sigma_{s}$ and 
$\sigma_{s}$, $\epsilon$, and $\mathcal{L}$ are the same-sign top pair signal cross section, the signal efficiency after cuts and detector effects, 
and the integrated luminosity, respectively.  The number of 
background events after all cuts is denoted by $n_{b}$. In Eq.\ref{poisson}, $\sigma_{s}$ is assumed to be a free parameter 
to be able to consider various FCNC signal cross sections. In order to find the upper limit at $95\%$ confidence level (CL) 
on the signal rate ($\sigma^{95\%}$), first one needs
to integrated over the posterior probability:
\beq \label{xx}
\frac{\int^{\sigma^{95\%}}_{0} \mathcal{P}(n | \epsilon\times\sigma_{s}\times \mathcal{L}, n_{b}) d\sigma_{s}}{\int^{\infty}_{0} \mathcal{P}(n | \epsilon\times\sigma_{s}\times \mathcal{L}, n_{b}) d\sigma_{s}} = 0.95,
\eeq
then solve the Eq.\ref{xx} under the assumption of $n = n_{b}$ after giving the inputs for the expected background, signal efficiency, and the integrated luminosity.
The signal efficiencies ($\epsilon$) for the three masses of $Z'$ are given in Table\ref{tab:eff}. 
The obtained bounds for various values of integrated luminosities are shown in Fig.\ref{fig:bounds}.  
As can be seen, the same-sign top quark pair enables us to constrain 
$a_{ut} \gtrsim 0.05 (0.03)$ for $m_{Z'} \sim 400 \ \gev$ and $a_{ut} \gtrsim 0.11 (0.065)$ for $m_{Z'} \sim 1$ TeV with
the integrated luminosity of  100 (3000) fb$^{-1}$. The limits on $a_{ct}$ are looser due to the lower cross section 
of processes involving charm initial states compared with that of up quarks: 
 $a_{ct} \gtrsim 0.11 (0.067)$ for $m_{Z'} \sim 400 \ \gev$, and $ a_{ct} \gtrsim 0.25 (0.1)$ for $m_{Z'} \sim 1 \ \tev$ 
 with 100 (3000) fb$^{-1}$ of data at $95\%$ CL.

Following exactly the same procedure as used for $Z'$ FCNC analysis and using
 the same cuts for the scalar mediator, the bounds presented in the bottom plots of Fig.~ \ref{fig:bounds} are derived. 
Due to the smaller cross section of the scalar, the bounds are weaker in terms of the scalar mass.
This is the case for both LHC limits and $D^0-\bar D^0$ limits.

\subsubsection{Transverse momentum based asymmetry }
\label{asy}

\begin{figure}[h!]
\centering 
\includegraphics[scale=0.65]{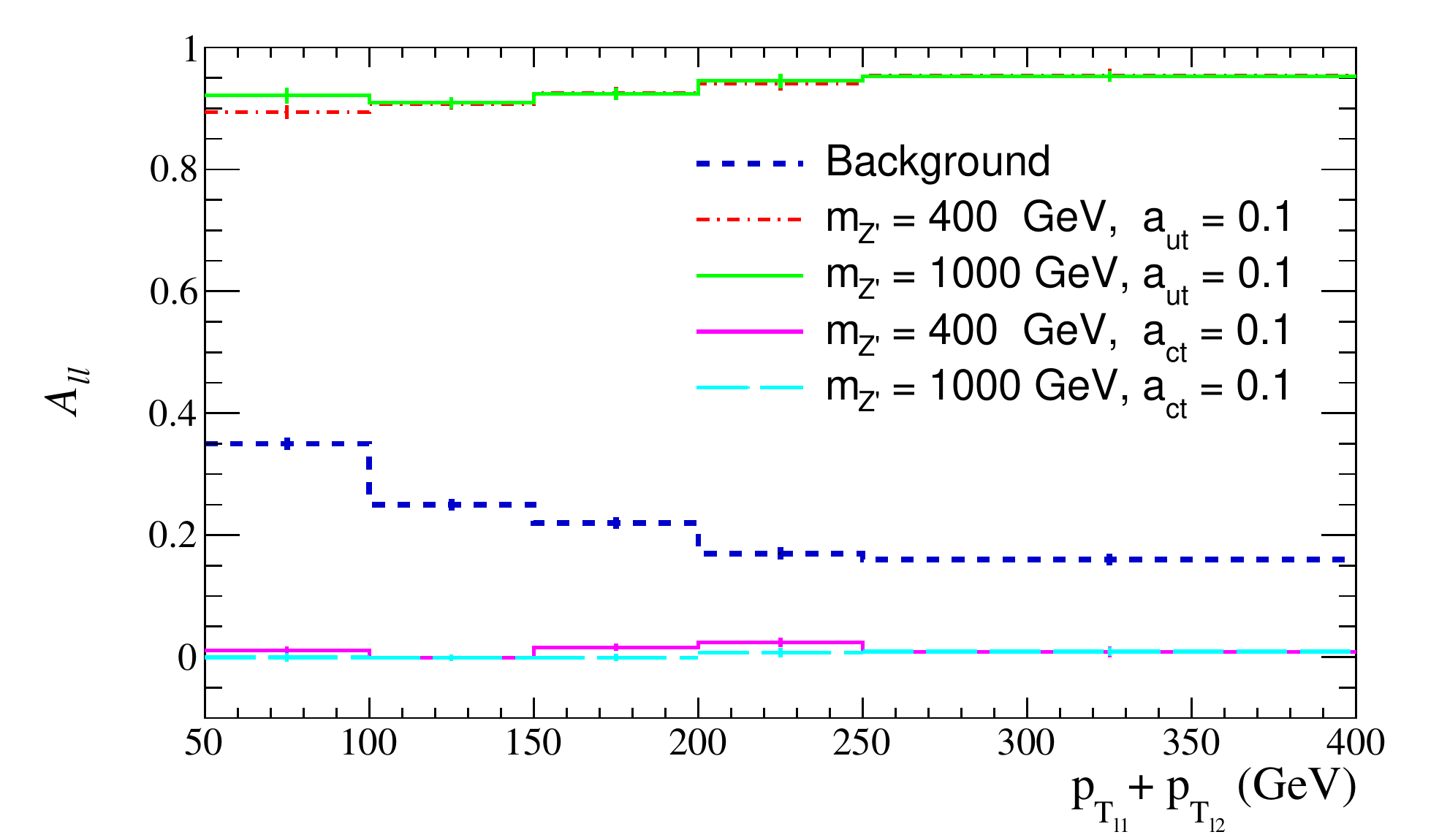}
\caption{The behavior of $A_{\ell \ell} (T)$ in terms of $T = p_{T\ell_{1}}+ p_{T\ell_{2}}$ for two benchmark points of signal and for the background.
 For the signal benchmarks with $a_{ut}$ turned on, the asymmetry has an upward trend due to
 the fact that up quark is a valence quark and anti-up quark is a sea quark. 
 For $a_{ct}$ turned on, both charm and anti-charm are sea quark leading to almost a constant and zero value for the asymmetry. }
\label{fig:asym}
\end{figure}

One of the striking characteristics of the signal,  $tt+\bar{t}\bar{t}$, 
is the asymmetry between the $tt$ and $\bar{t}\bar{t}$ cross sections.
The production rates of $uu \to tt$ and $\bar{u}\bar{u} \to \bar{t}\bar{t}$ are different at the LHC 
due to the fact that $u$-quark is a valence quark and carries larger fraction of proton momentum ($x$) with respect
to the $\bar{u}$-quark which is a sea quark with a PDF peak at low energies.
This feature is quite helpful in discriminating the signal from the background and has been already 
proposed in Ref.\cite{Durieux:2012gj} to study the baryon and lepton number violation at the LHC. 
As the parton distribution functions of $c$-quark and $\bar{c}$ quark are the same, 
the cross sections of $cc \to tt$ and $\bar{c}\bar{c} \to \bar{t}\bar{t}$ are expected to be similar.
In the same-sign dilepton decay channel, the $tt$/$\bar{t}\bar{t}$ asymmetry can be directly observed 
in the charges of same-sign dilepton. This assumption is realistic as the lepton selection efficiencies and 
lepton contaminations from fake do not depend on the charge.  As a result, to distinguish between the
$tuX$ and $tcX$ signal scenarios and to separate the FCNC signal from the SM backgrounds,  
we can define a momentum based charge asymmetry in the following form:
\beq
A_{\ell \ell}  (T) \equiv \frac{N_{\ell^+ \ell^+} (T) - N_{\ell^- \ell^-}(T )}{N_{\ell^+ \ell^+} (T)+ N_{\ell^- \ell^-}(T)}, \text{where}~T = p_{T\ell_{1}}+ p_{T\ell_{2}},
\label{eq:asymmetry}
\eeq
where $p_{T}(\ell)$ is the lepton transverse momentum and $N_{\ell^+ \ell^+}(N_{\ell^- \ell^-})$ is the number of events
with opposite(negative)-sign dilepton.  
The distribution of the $A_{\ell \ell}  (T)$ is presented for both FCNC signal scenarios $tuZ'$
and $tcZ'$ and for the main background process in Fig.\ref{fig:asym}. The cross section for 
the same-sign dilepton decay channel of the main background process {\it i.e.} $t\bar{t}W^{\pm}$ at the next-to-leading order
has been calculated in Ref.\cite{Campbell:2012dh}.
For high $p_T$ values, the number of positive leptons should be much higher than the negative leptons. 
Hence, if we look at this asymmetry for different intervals of $T$,  an upward trend is observed.
This is while, in the background the asymmetry has a downward trend versus $T $. 
That is because as shown in Fig.~\ref{fig:bkg}, the initial states that lead to $\ell^+ \ell^+$, can be both valence quarks, 
one valence quark and one sea quark, or both sea quarks. Similarly, the ones that will make $ \ell^- \ell^-$ 
can also be any combinations of valence quark and sea quark. There is a slight preference for positive
 leptons $ \ell^+\ell^+$, due to the slightly higher PDF of $u$-quarks compared with $d$-quarks, leading to a small positive $A_{\ell \ell}$. 
 High $p_T$ limit corresponds to high $Q^2$ in the PDFs, and in high $Q^2$, diagrams with sea quark initial state 
 will contribute more significantly, and thus cause the asymmetry to fade off. Furthermore, the difference between the up 
 and down PDF in high $Q^2$ limit also becomes more negligible which is another reason the asymmetry in the background has a modest decrease towards zero. 

\begin{figure}[h!]
\centering 
\includegraphics[scale=0.4]{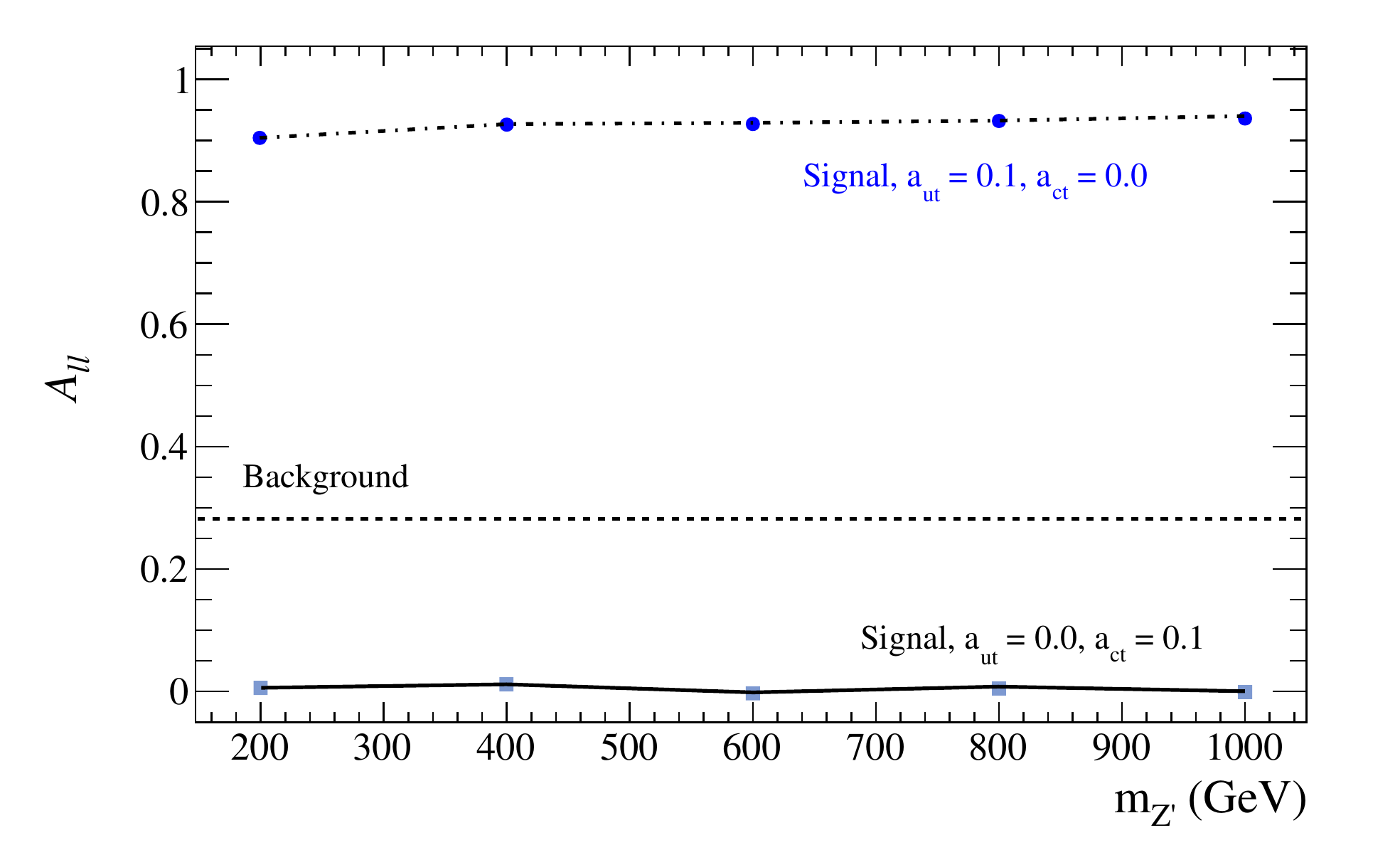}
\includegraphics[scale=0.4]{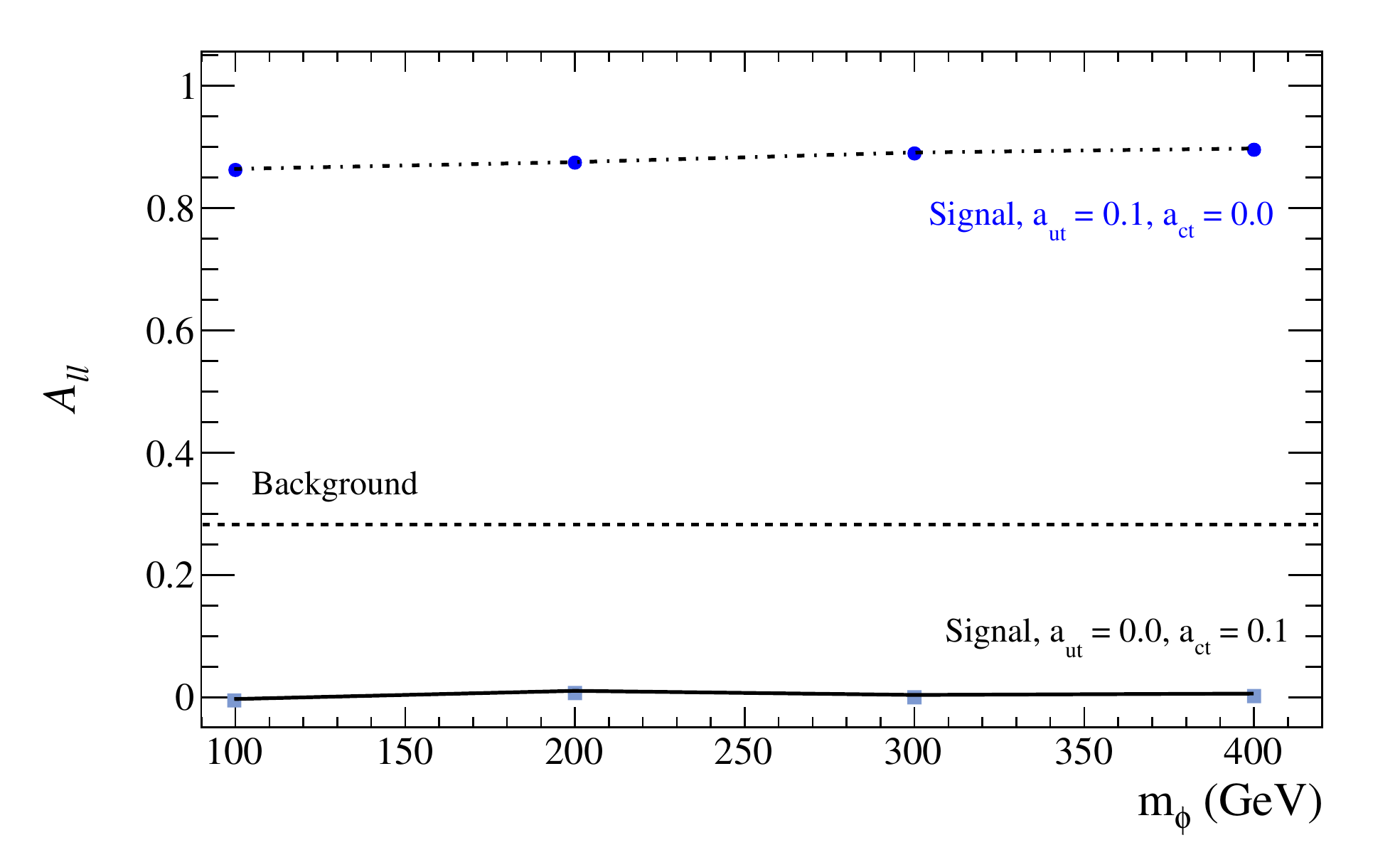}
\caption{The total asymmetry is shown as a function of $m_{\text{med}}$. Note that for the case of
 $a_{ut} \neq 0$ and $a_{ct} = 0$, the asymmetry increases as a function of $m_{Z'/\phi}$. 
 For $a_{ct} \neq 0$ and $a_{ut} = 0$, however, the total asymmetry is zero for all masses of $Z'$ and $\phi$.  }
\label{fig:tot_asy}
\end{figure}

We present the total asymmetry (summed over all bins) 
for various $m_{Z'}$  in Fig.~\ref{fig:tot_asy}, which demonstrates a small increase in the value of the asymmetry 
for larger $m_{Z'}$.  For the case that we turn $a_{ut}$ off, and let $a_{ct}$ take non-zero values, since charm and anti-charm 
are both sea quarks with almost the same PDF behavior, the asymmetry is almost constant at zero. This result is consistent for all $m_{Z'}$ tested.

It is important to mention that the introduced asymmetry in Eq.\ref{eq:asymmetry} is sensitive to the choice of proton parton distribution
functions (PDFs). In a measurement for such an observable ($A_{\ell\ell}$), one needs to include a source of systematic due to the limited knowledge of the proton PDF.
For instance, we estimated an uncertainty from the choice of PDF by calculating the total asymmetry with three PDF sets: CTEQ6.6 \cite{Pumplin:2002vw}, 
NNPDF \cite{Ball:2012cx}, and MSTW \cite{Martin:2009iq}.
The maximum relative variation on the asymmetry is found to be $\Delta A_{\ell\ell}/A_{\ell\ell} = 0.5\%$ for the vector FCNC couplings with $a_{ut} = 0.1$ and $m_{Z'} = 1$ TeV.

\subsubsection{Angular observables }
\label{kurt}
\begin{figure}[htb]
\centering 
\includegraphics[width=0.4\textwidth, height=0.21\textheight]{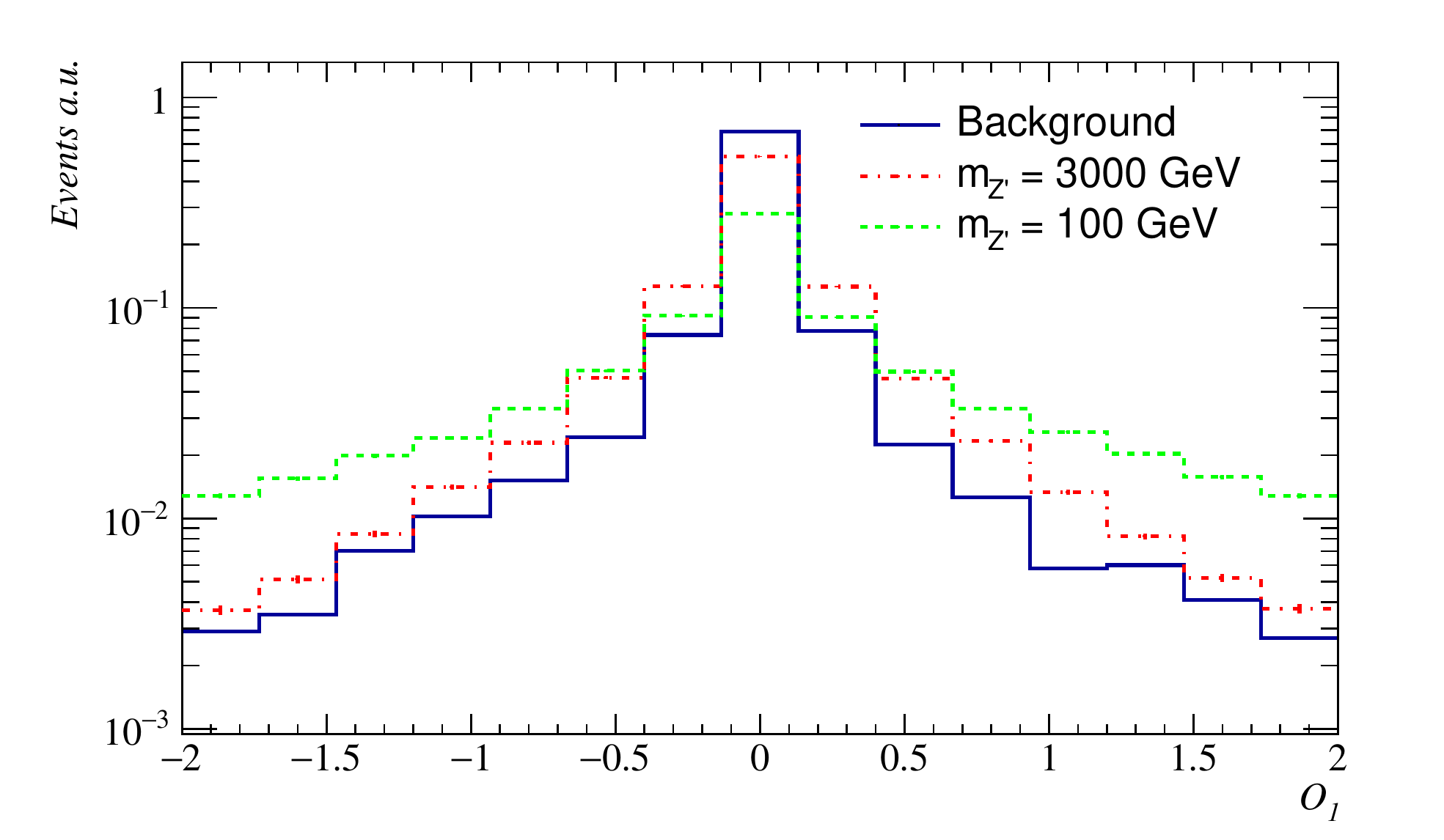}
\includegraphics[width=0.4\textwidth, height=0.21\textheight]{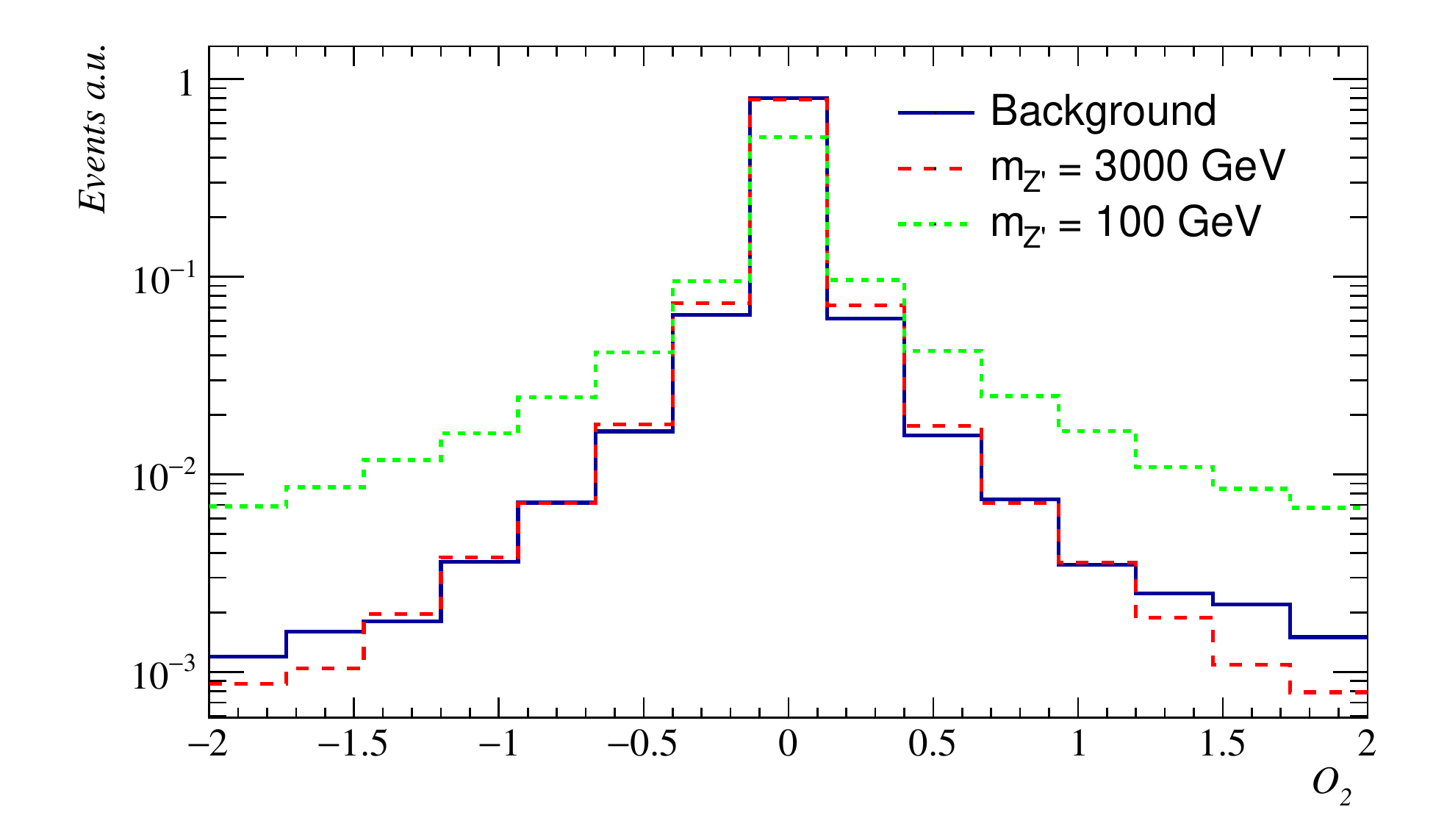}
\caption{The distributions of $O_1$ and $O_2$ (defined in Eq.~\ref{planes}) for the $tt$ signal with the $Z'$ masses of 100 GeV and 3000 GeV and
for the SM background.}
\label{fig:kurto102}
\end{figure}

In the last subsection, it has been shown that the transverse momentum based asymmetry $A_{\ell\ell}$
is sensitive to $tqX$ FCNC couplings enabling us to distinguish between the $tuX$ and $tcX$
interactions. However, $A_{\ell\ell}$ has shown negligible sensitivity to the mass of the mediator $m_{Z'/\phi}$. 
In this subsection, the concentration is on introducing sensitive observables to the mass of the FCNC mediator
and could discriminate between the scalar and vector FCNC couplings and the background.
In this study, one main difference between the scalar and vector signal scenarios and the background is that in the background process
there is no specific correlation between the two jets or the two leptons while for the signal processes, the final state particles are highly correlated. 
This feature of the signal processes (scalar and vector) and background could be examined using the 
 the four momenta of the visible particles in the detector. We look at the following variables:
\begin{align}
O_1 &= \frac{ [\hat z \cdot (\vec{p}_{b_{1}} \times \vec{p}_{b_{2}} )] [\hat z \cdot (\vec{p}_{b_{1}} - \vec{p}_{b_{2}})]}{m_t^3} \nonumber\\
O_2 &= \frac{ [\hat z \cdot (\vec{p}_{\ell_{1}} \times \vec{p}_{\ell_{2}} )] [\hat z \cdot (\vec{p}_{\ell_{1}} - \vec{p}_{\ell_{2}})]}{m_t^3}
\label{planes}
\end{align}
where $b_{_1}\ (\ell_{_1})$ is the leading jet (lepton), and $b_{_2}\ (\ell_{_2})$ is the second leading jet (lepton). 
The momentum vector of each object is specified by $p$, and $\hat z$ is the direction along the beam.
The distributions of $O_{1}$ and $O_{2}$ observables for the $tuZ'$ signal with $m_{Z'} = 100, 3000$ GeV and for 
the background are depicted in Fig.\ref{fig:kurto102}. As it can be seen, the distributions peak at zero and 
the $O_{1}$ and $O_{2}$ distributions become narrower with a sharper peak at zero when the mass of $Z'$ increases.
The behaviors of the $O_{1}$ and $O_{2}$ could be understood by looking again at the
squared matrix element of the signal process presented in Fig.\ref{me}.
At low values of $\hat{s}$, the squared matrix element 
tends to small values and in particular for larger $m_{Z'}$, it grows with $\hat{s}$. 
As a result, the the top quarks in the for large $Z'$ mass are  boosted which means their decay products are almost colinear, and almost along 
the same direction. Since  the top quarks are nearly back-to-back, the jets and leptons are expected to be back-to-back as well. 
Hence, $p^{b(\ell)_{1}} \times p^{b(\ell)_{2}} \rightarrow 0$, leading to $O_{1}=0$ and $O_{2}=0$. 
Thereby, the $O_{1,2}$ distributions for the signal processes should have a peak in around zero
, and the peak gets sharper as the mass of $Z'$ increases.

\begin{figure}[htb]
\centering 
\includegraphics[width=0.8\textwidth, height=0.3\textheight]{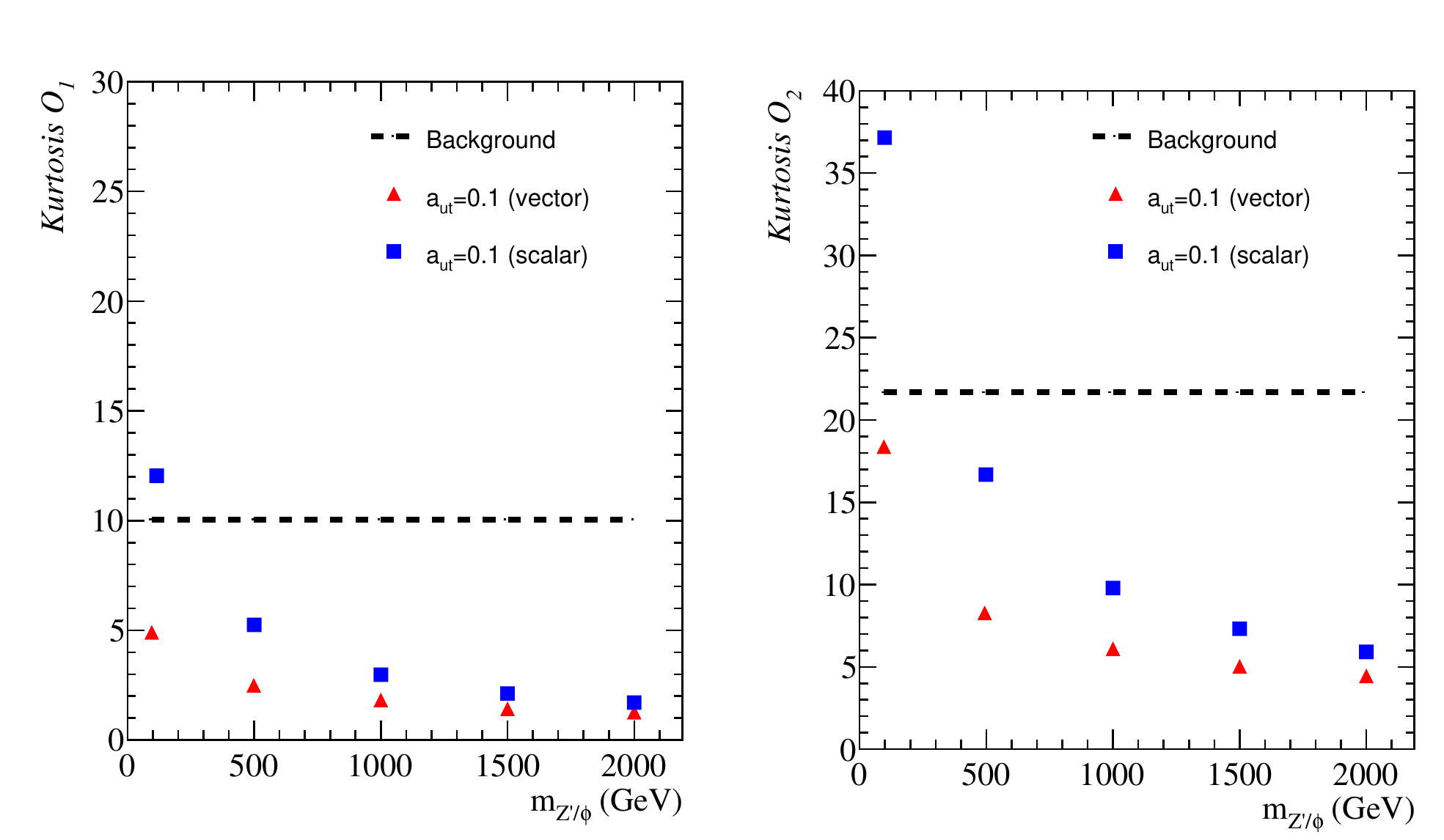}
\caption{The kurtosis of our benchmark points in the $O_1$ and $O_2$ (defined in Eq.~\ref{planes}) distribution. 
The black solid line belongs to the SM value. The values that belong to to the 
$Z'$ mediation are shown  in blue, and the ones with scalar mediator $\phi$,  is in red. 
The Kurtosis for lighter mediators is higher than the heavier ones. 
Moreover, the peaks when $\phi$  is the mediator is consistently sharper than when $Z'$ mediates.}
\label{fig:O1O2}
\end{figure}

A measure which quantifies the sharpness of the peak and heaviness of the tail is
\textit{kurtosis}. Sharper peaks lead to smaller value of kurtosis. We can then compare the kurtosis for various values of $m_{Z'/\phi}$ 
 with respect to the SM. Fig.~\ref{fig:O1O2} demonstrates the values of kurtosis for few benchmark points and the black dashed-line corresponds to the SM background value. 
The argument presented solely depends on the value of $\sqrt{\hat s}$. To see how the peaks at zero varies for different benchmarks, 
recall that for larger mediator mass, the effect of $ \sqrt{\hat s}$ is more significant. Consequently, the peak should be sharper for heavier $Z'$ mass, as shown in Fig.~\ref{fig:kurto102}.

The $O_1$ and $O_2$ variables are able to distinguish between $\phi$ and $Z'$ mediators as they have 
different spins which leads to some difference in the angular 
distribution of the final top quarks.  As shown in Fig.~\ref{fig:O1O2}, the difference is very noticeable especially for smaller $m_{Z'/\phi}$. 

\subsection{Same-Sign top pair plus a $W$ boson }
\label{sec:ttW}

\begin{figure}[h!]
\centering 
\includegraphics[width=0.35\textwidth, height=0.2\textheight]{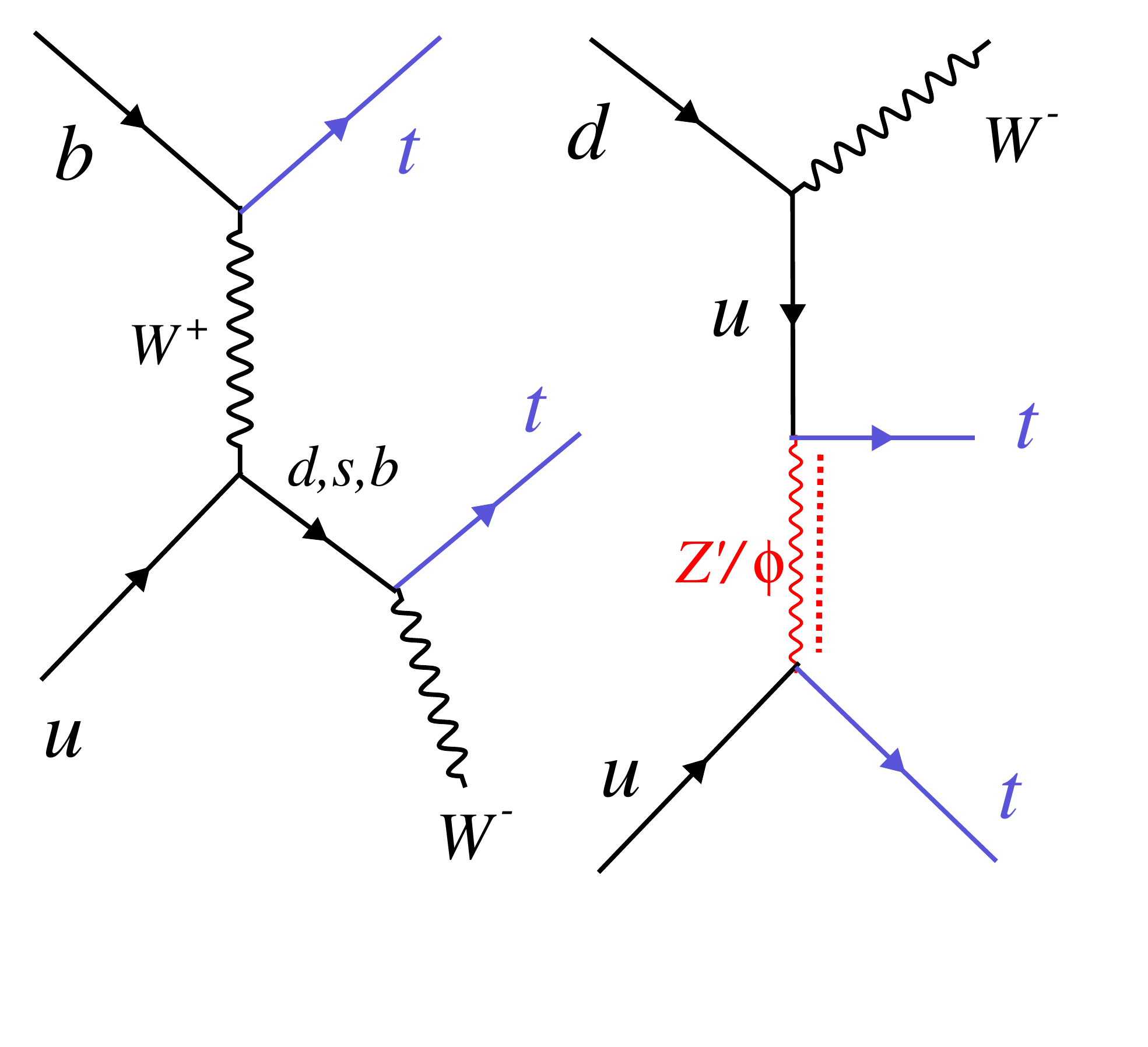}
\caption{Representative Feynman diagrams for production of same-sign top quark pair in association with a $W$ boson in the SM (left) and
in the vector and scalar FCNC model (right) at the LHC. }
\label{fig:ttw-feynman}
\end{figure}

To further enhance the sensitivity to the $tqZ'$ and $tq\phi$ flavor changing, 
we also look at same-sign top pair + $X$, where $X$ could be either a gauge boson or a jet. 
$X$ can be determined such that the ratio of signal over background is the most optimal or a good sensitivity is achieved.
For the cases $X =$ jets, photon, or $Z$, since there are many colored or electromagnetic charged intermediate states in the background, 
the diagrams of the background increase much faster than the one of the signal.
In particular, for $X= $ jet, a significant increase in the signal cross section is observed however the
background  is more troublesome. 
For $X=W$, the increase in the diagrams of the background is more tame, and a better chance of improving the sensitivity
is expected.  The representative leading order Feynman diagrams
for $ttW$ in the SM (left) and in the pure FCNC model (right) are presented in Fig.\ref{fig:ttw-feynman}.

The cross section of the $ttW$ in the presence of the FCNC couplings  is expected to be smaller than the $tt$ signal because 
we are exchanging one of the up quarks with a down quark in the initial state and also $ttW$ process has a smaller 
phase space due to the presence of a heavy $W$ boson in the final state. 
The production cross section of the $ttW$ process  versus the 
mass of $Z'$ and $\phi$ is presented in Fig.\ref{xsec}.
Less amount of background than the $tt$ process would contribute
which makes this channel interesting enough to study.
To still benefit from same-sign top pair features, the leptonic decay of the top quarks is considered and the hadronic decays of the
$W$ boson is taken into account as it has higher branching fraction than the leptonic.
The main background processes are $t\bar{t}V$ with $V = W, Z$, $W^\pm W^\pm +$jets and four top quarks production.
The irreducible SM $ttW$ production shown in the left side of Fig.\ref{fig:ttw-feynman} is expected to 
be negligible due to CKM suppression. Consequently, it is not considered in the analysis.

The same as the same-sign top study, the signal and background events are generated with {\tt MadGraph5-aMC@NLO}. Then the generated events 
are passed through {\tt Pythia 6} for showering and hadronization, and finally to {\tt Delphes 3} to inlcude the detector level effects. 
The signal events are selected by requiring exactly two same-sign charged leptons with $p_T^{\ell} > 25 \ \gev$ and
$|\eta| < 2.5$. The charged leptons are required to be well-isolated with RelIso $ < 0.15$, where RelIso is defined in Eq.\ref{iso}.
Jets are reconstructed using the anti-$k_{T}$ algorithm with a distance parameter of 0.4.
Each event has to have at least four jets with $p_T(j)> 30 \ \gev$ and $|\eta_{j}| < 2.5$ from which 
two should be $b$-tagged.  The missing transverse energy $E^{miss}_{T}$  is required to be greater than 40 GeV. For further background reduction, a lower cut is applied on the transverse mass
of the final state defined as:
\begin{eqnarray}
m_{T} = \sqrt{(E_{Tb1}+E_{Tb2}+E_{T\ell1}+E_{T\ell2} + E^{miss}_{T})^{2} - (\vec{p}_{Tb1}+\vec{p}_{Tb2}+\vec{p}_{T\ell1}+\vec{p}_{T\ell1}+\vec{p}^{miss}_{T})^{2}},
\label{masst}
\end{eqnarray}

\begin{figure}[h!]
\centering 
\includegraphics[width=0.55\textwidth, height=0.25\textheight]{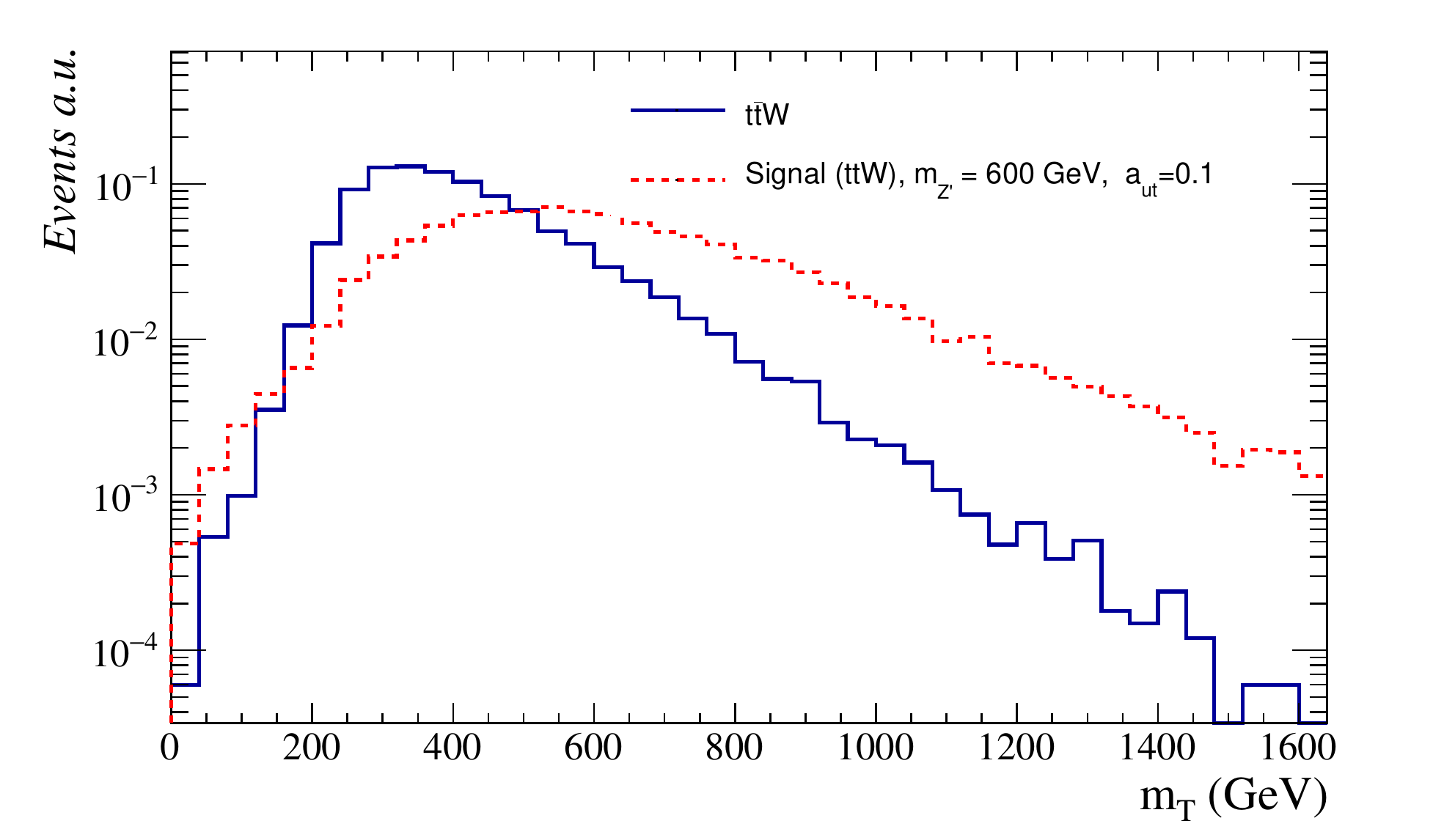}
\caption{The distribution  of the transverse mass $m_{T}$ of the system as defined in Eq.\ref{masst} for the signal with $a_{ut} = 0.1, m_{Z'} = 600$ GeV 
and  for the $t\bar{t}W$ background process.}
\label{me11}
\end{figure}

The distribution of the $m_{T}$ for the $ttW$ signal with $m_{Z'} = 600$ GeV and $a_{ut} = 0.1$ and for the
the main background process {\it i.e.} $t\bar{t}W^{\pm}$ are presented in Fig.\ref{me11}. As it can be seen, the
signal events tend to have a peak at larger value with respect to the background. For the background, 
the peak is around 300 GeV while the signal peaks at around 550 GeV.  Therefore, applying a lower cut 
on $m_{T}$ is useful to suppress the background contribution considerably. A minimum value of $350$ GeV
is applied on $m_{T}$.

After the cuts, sum of the  background cross sections
is found to be $0.28 \ \text{fb}$. The signal efficiencies 
after the cuts for the $Z'$ masses of 400 GeV, 600 GeV, 1 TeV are $2.5\%$, $2.6\%$, $2.5\%$, respectively.

\begin{figure}[h!]
\centering 
\includegraphics[scale=0.35]{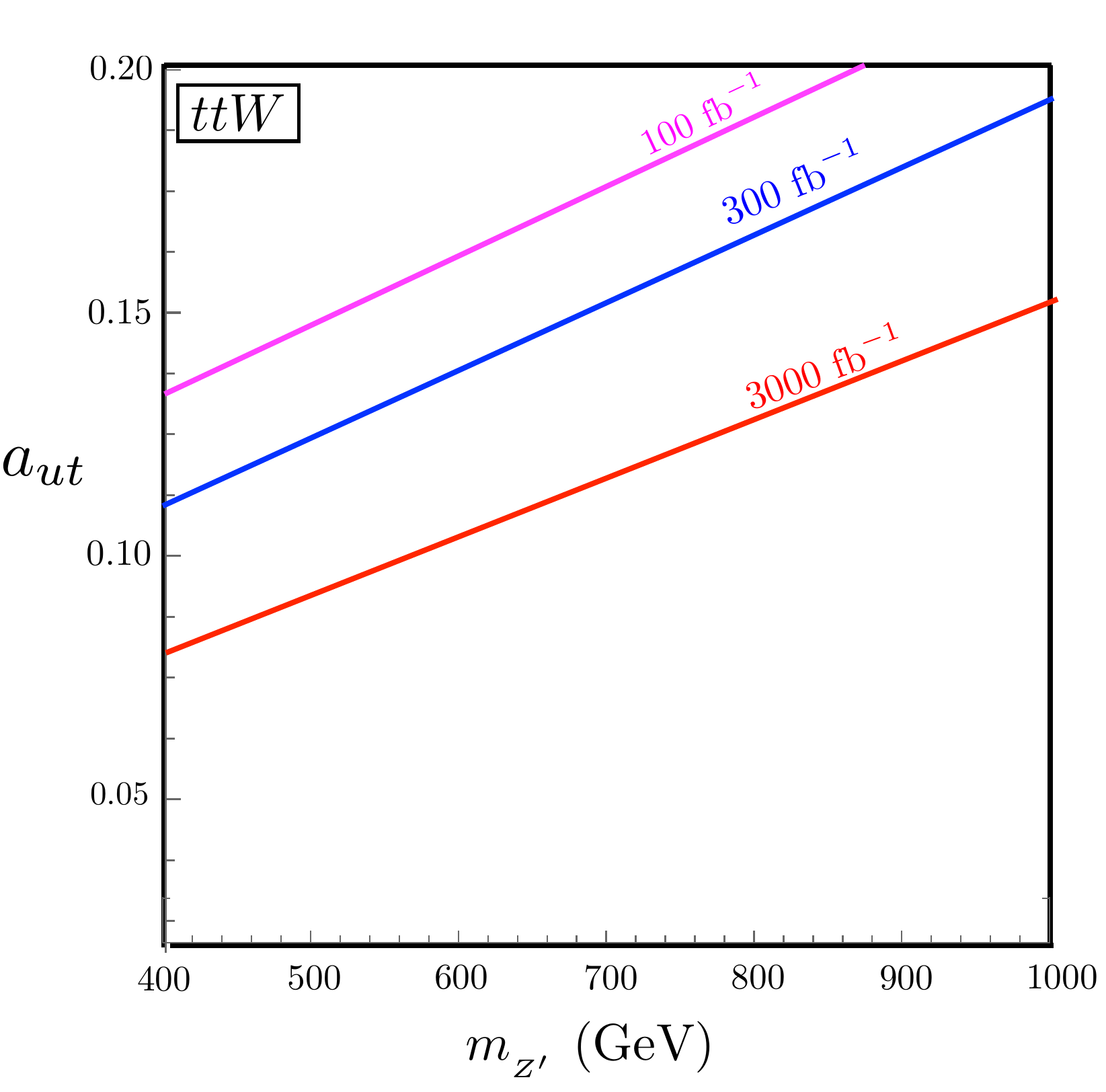}
\includegraphics[scale=0.35]{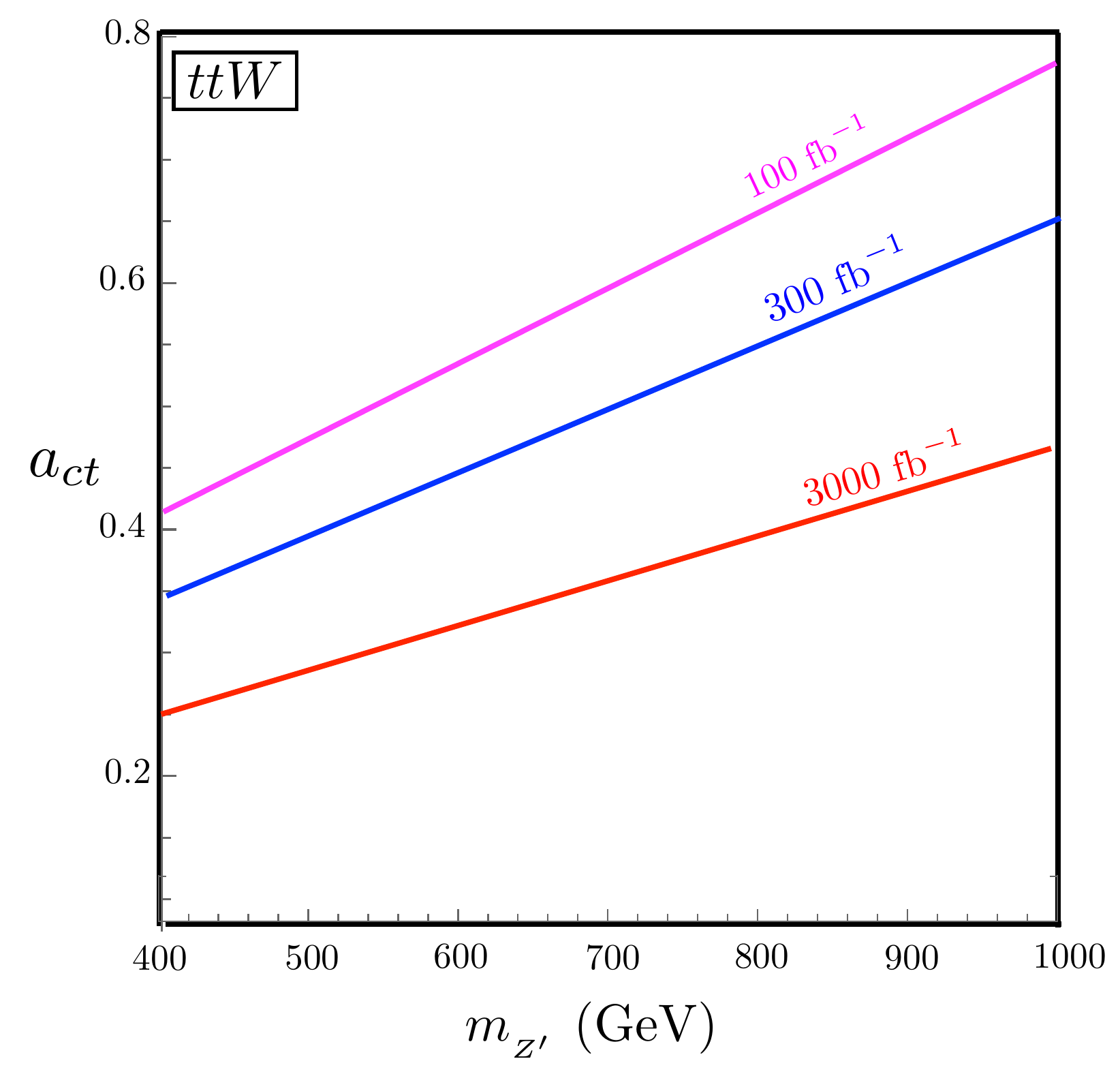}\\
\includegraphics[scale=0.35]{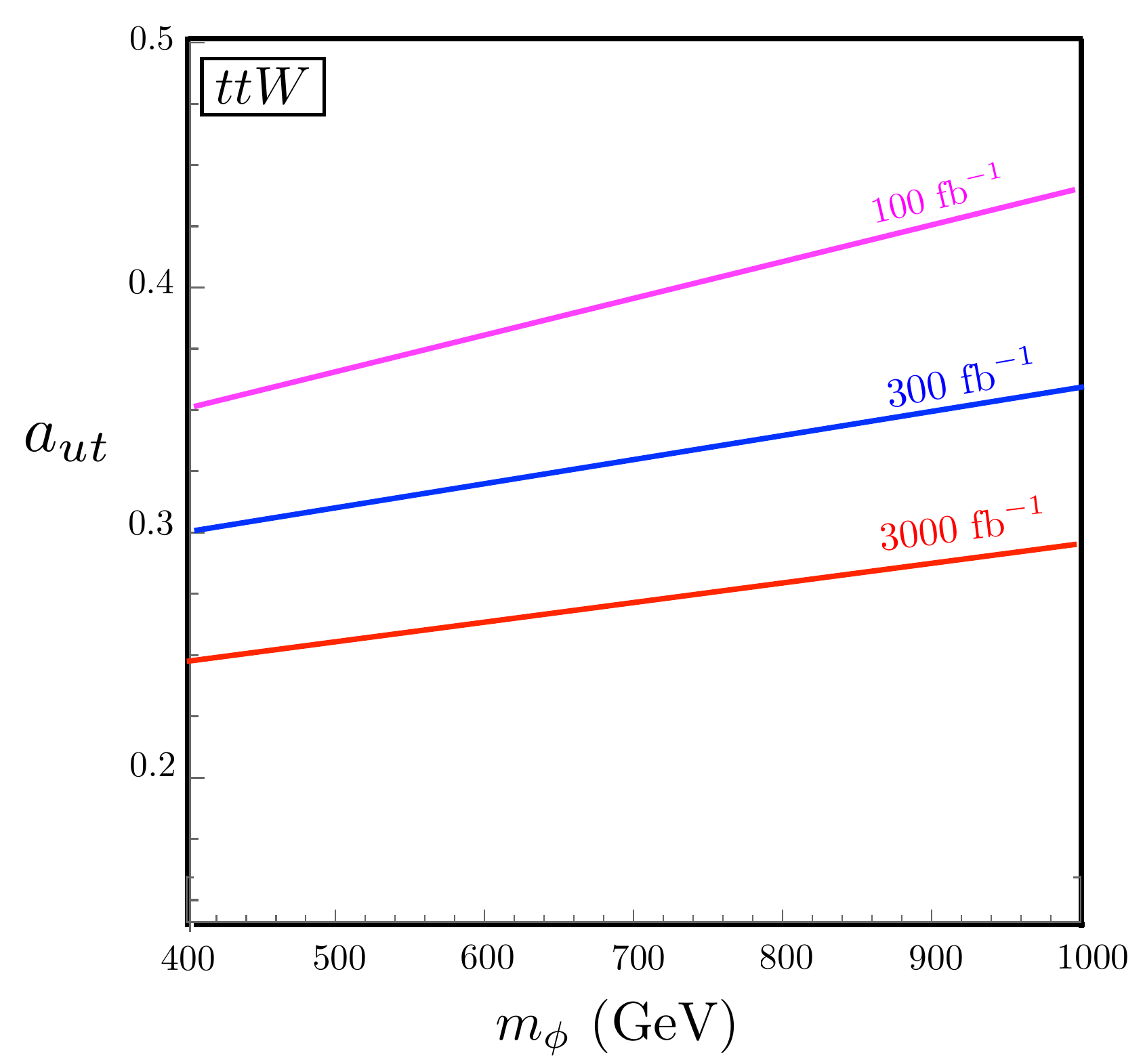}
\includegraphics[scale=0.35]{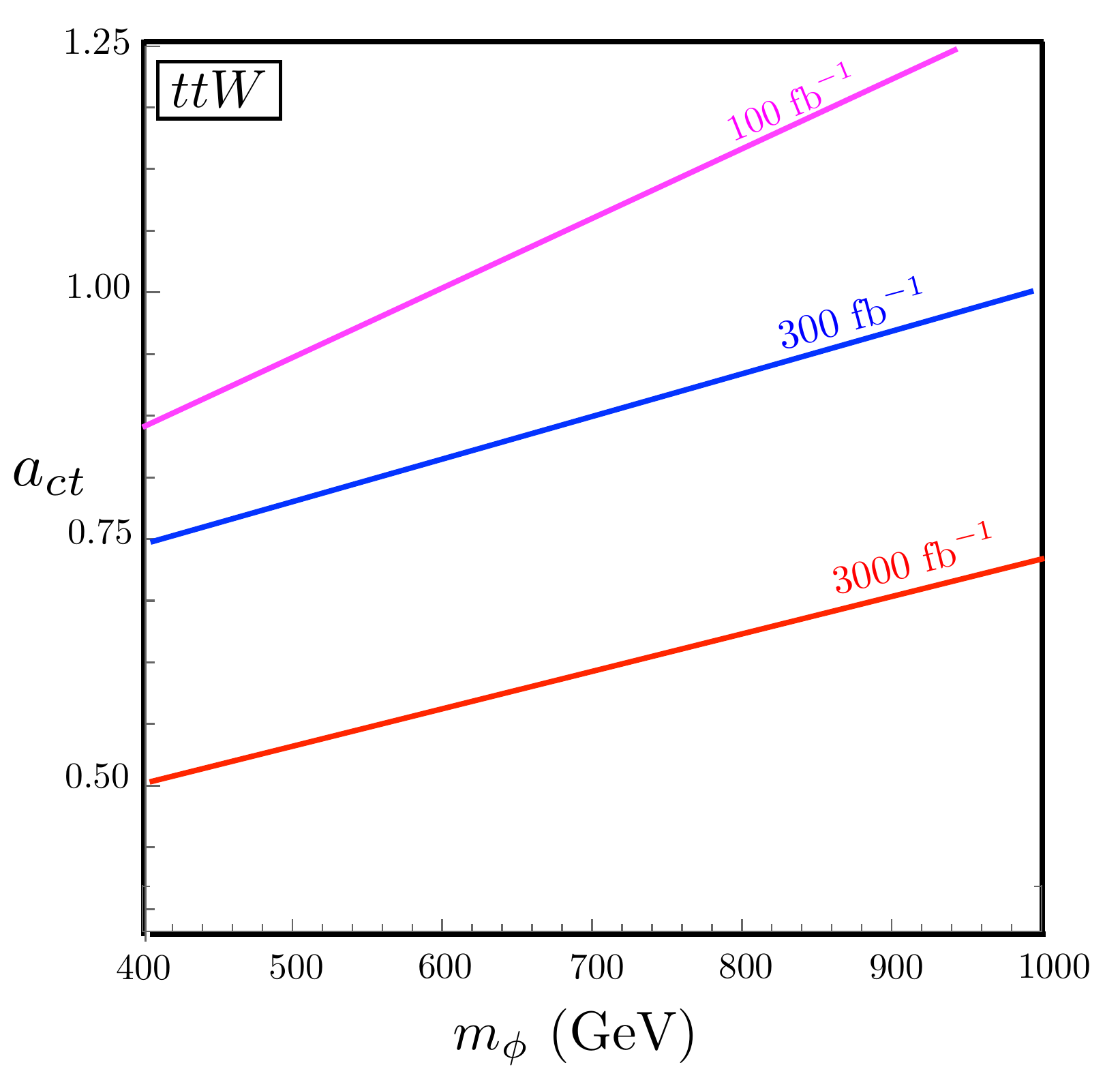}\\
\caption{The $95\%$ CL limits on $Z'$ (up) and $\phi$ (down) that have flavor changing coupling with up and top quarks (left) and with charm and top quarks (right). 
The exclusion regions are presented  with $ 100 \ \fb^{-1}$ (pink), $300 \ \fb^{-1}$ (blue) 
and $3000 \ \fb^{-1}$ (red) integrated luminosity.  }
\label{fig:boundsttw}
\end{figure}

Using the same statistical procedure as the one explained previously in Eq.\ref{xx}, upper limits
at the $95\%$ CL are set on the $ttW$ cross section and the limits are translated into constraints
on the parameter space of the model, \textit{i.e.} $(a_{qt},m_{Z'/\phi})$.
Figure \ref{fig:boundsttw} shows the bounds  at $95\%$ CL on vector flavor changing and scalar flavor changing 
with up and top quarks and with charm and top quarks. 
The exclusion regions are depicted  with $ 100 \ \fb^{-1}$, $300 \ \fb^{-1}$ 
and $3000 \ \fb^{-1}$ integrated luminosities.

According to Fig.\ref{fig:boundsttw}, the $ttW$ process excludes the parameters  
$a_{ut} \gtrsim 0.13 (0.08)$ for $m_{Z'} \sim 400 \ \gev$ and $a_{ut} \gtrsim 0.21 (0.14)$ for $m_{Z'} \sim 1$ TeV with
 100 (3000) fb$^{-1}$ of  integrated luminosity.  As expected, the limits on $a_{ct}$ is looser 
which is found to be $a_{ct} \gtrsim 0.42 (0.25)$ for $m_{Z'} \sim 400 \ \gev$, and $ a_{ct} \gtrsim 0.77 (0.42)$ for $m_{Z'} \sim 1 \ \tev$
for the integrated luminosity of 100 (3000) fb$^{-1}$.
Going to higher integrated luminosities will improve the sensitivity on $a_{ut}$ and $a_{ct}$  by a factor 
of around $\lesssim 2$. For the scalar scenario, the strongest limit on the $a_{ut}$ and $a_{ct}$ are 0.15 and 0.45, respectively.
The exclusion regions obtained from the $ttW$ process are looser with respect to those derived from the $tt$ process. However,
a comparison of limits shows that, for a given integrated luminosity, the exclusion borders from $tt$ channel have larger slopes in $tt$ process
than the $ttW$.  As a result, the combination 
of these two channels would improve the exclusions limits in particular for the large $Z'/\phi$ mass regions. In the next subsection, the
combination procedure and results will be presented.

\subsection{Combination of the $tt$ and $ttW$ channels }
\label{sec:comb}
So far, the sensitivities of the same-sign top and same-sign top associated with 
a $W$ boson have been studied and the exclusion limits presented in the previous sections.
Here, the goal is to combine the two analyses which is expected to provide a better sensitivity 
to the parameter space of the model.

In order to derive the bounds from the combination of  the
two production mechanisms $tt$ and $ttW$, the same statistical  technique as explained in Eq.\ref{poisson}  and Eq.\ref{xx}
is used. The number of signal is defined as $n_{s} = \sum_{i=channels} \epsilon_{i} \times \sigma_{i} \times \mathcal{L}$, where
$i$ runs over the contributing processes.
It should be noted that the cut efficiencies $ \epsilon_{i}$ vary depending on the production
mechanisms and also for each process the efficiency is dependent on the mass of the $Z'/\phi$.
Therefore, to derive $95\%$ CL exclusion limits on the model parameters $a_{qt}$ and $m_{Z'/\phi}$,
one needs to properly considers the selection efficiencies.

\begin{figure}[h!]
	\centering 
	\includegraphics[width=0.42\textwidth, height=0.3\textheight]{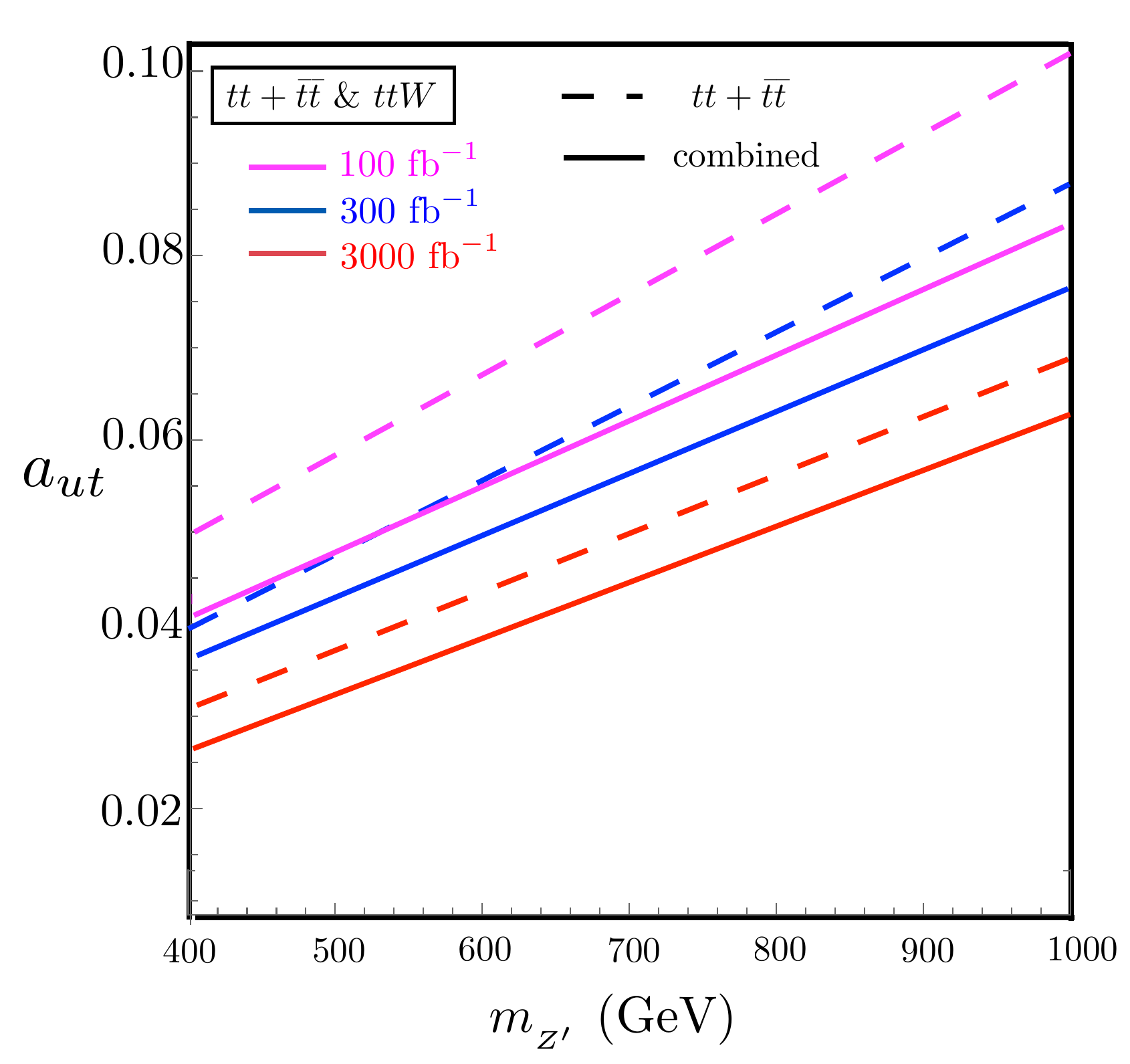}
		\includegraphics[width=0.42\textwidth, height=0.3\textheight]{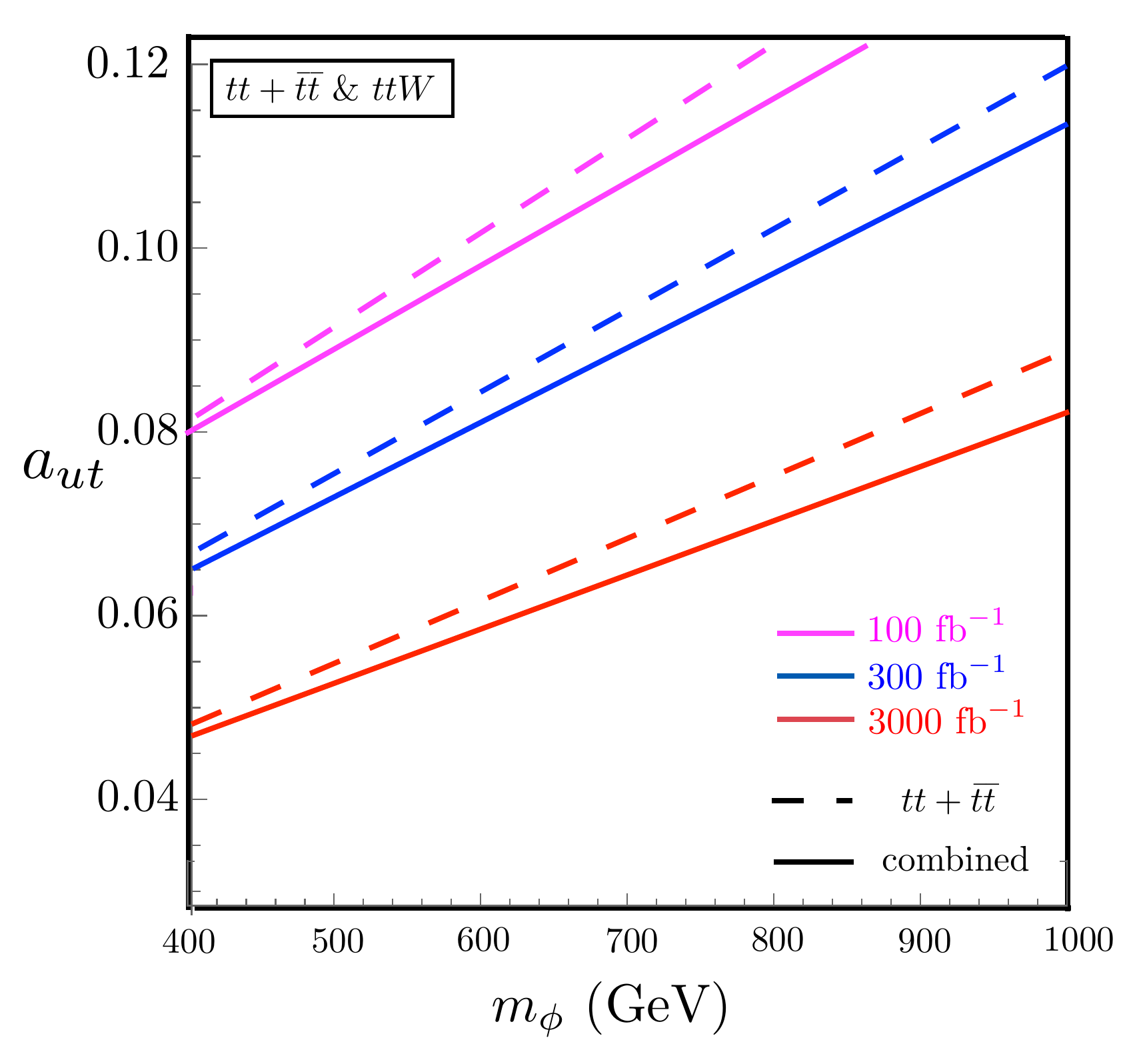}
	\caption{The limits on $Z'$ that has flavor changing coupling with up and top quarks. 
	The rest of the flavor changing couplings are set to zero. 
	The excluded regions for $tt$ (dashed) and combination of $tt$ and $ttW$ (solid) is shown. 
	The regions to the up of the plots are excluded in 95\% CL for $ 100 \ \fb^{-1}$ (magenta), $300 \ \fb^{-1}$ (blue) and $3000 \ \fb^{-1}$ (red) integrated luminosity.}
	\label{fig:combinedvmet}
\end{figure}

The exclusion plots of the combined channels at $95\%$ CL for 
$(a_{ut}, m_{Z'})$ and  $(a_{ut}, m_{\phi})$ are depicted in Fig.\ref{fig:combinedvmet}.
The combination improves the upper limit on  $a_{ut}$ for various masses of $Z'$ and $\phi$ in particular  with 
the integrated luminosities of 100  fb$^{-1}$ and 300 fb$^{-1}$. With an integrated luminosity of 100 fb$^{-1}$,
for $m_{Z'} (m_{\phi})= 1$ TeV, the upper limit on $a_{ut}$ gets improved with an amount of  $20\% (30\%)$.

In Refs.\cite{CMS:2016wdk,Aad:2014wza }, the ATLAS and CMS experiments presented the results of searches for monotop events using
the proton-proton collisions at the center-of-mass energy of 8 TeV with an integrated luminosity of  20.3 fb$^{-1}$ and 19.7 fb$^{-1}$, respectively.
Both searches are based on the leptonic decays of the top quark which is produced associated  with missing energy.
For a coupling strength of $a_{qt} = 0.1$, ATLAS and CMS excluded  $Z'$ mass lower than 523 GeV and 432 GeV, respectively.
From the the combined analysis of $tt$ and $ttW$ channels, with 100 fb$^{-1}$, a lower limit of $\sim1500$ GeV is obtained for $a_{ut} = 0.1$.

\section{Summary and conclusions}
\label{sec:conclusions}

In this paper, we have performed a detailed analysis to 
search for flavor violation effects in the top quark sector 
following a simplified theory approach. 
In the considered model, the tree-level 
couplings of $tqZ'$ and $tq\phi$ with $q=u,c$ are
allowed.   Such FCNC couplings allow the productions of 
same-sign top ($tt+\bar{t}\bar{t}$) and same-sign top in association with 
a $W$ boson ($ttW^{-}+\bar{t}\bar{t}W^{+}$) 
which are complementary channels besides the monotop signature.
The total cross sections of the $tt$ and $ttW$ processes are
around 3000 fb and 200 fb for $a_{ut} = 0.25$ and $m_{Z'} = 1$ TeV. 
We concentrate on the same-sign dilepton channel for the $tt$ process, where both top quarks decay leptonically and
for the $ttW$, the same-sign dilepton decays of the top quarks and the hadronic decay of the $W$ boson are considered.
Same-sign dilepton events are a striking sign of physics beyond the SM at the LHC.
The analyses have been done by taking into account 
the response of a CMS-like detector and the contributions of the main background processes.
For both channels,  sets of kinematic variables
have been proposed to discriminate the signal events from the main background processes. 
The $95\%$ CL exclusion regions on the FCNC coupling strengths versus $m_{Z'/\phi}$ have been obtained  
with different scenarios of the integrated luminosities of 100 fb$^{-1}$, 300 fb$^{-1}$, and 3000 fb$^{-1}$.
We also have proposed a momentum dependent charge asymmetry 
as a powerful tool to discriminate between $tt$ signal and background
which also has the ability to separate the $tuX$ signal scenario from $tcX$, where $X=Z',\phi$.
It has been found that the combination of the two processes $tt$ and $ttW$ would 
improve the constraints on the flavor changing coupling $a_{qt}$ between $10\%$ to $20\%$.
For $m_{Z'} (m_{\phi}) = 1$ TeV, any value of the flavor changing coupling
above $a_{ut} \gtrsim 0.06 (0.08)$ is excluded with 3000 fb$^{-1}$
at $95\%$ CL.

\vspace{0.5 cm}
{\bf Acknowledgements:}\\
M. Mohammadi Najafabadi would like to acknowledge INSF for the support and S.M. Etesami for the 
useful discussions. F. Elahi is grateful to S. Tizchang for insightful discussions.
\\

\appendix
\section{Squared Matrix Element of the Same-Sign Top }
\label{app:xsection}

The squared matrix element of $u u \rightarrow t t$ for a flavor changing scalar mediator $\phi$ is  

\begin{equation}
\begin{split}
   \overline{\abs{\mathcal{M}}}_{uu\rightarrow tt}^2 &= \frac{(m_t^2-t)^2}{(m_\phi^2-t)^2} +\frac{(m_t^2-u)^2}{(m_\phi^2-u)^2}\\
    &+\frac{2 m_t^4-2 m_t^2 (-2 s+t+u)-s^2+t^2+u^2}{2 (m_\phi^2-t) (m_\phi^2-u)}
\end{split}
\end{equation}
where $s,t,u$ are the Mandelstam variables. 
The squared matrix element  of $u u \rightarrow t t$ for a flavor changing vector mediator $Z'$ is  

\begin{equation}
    \begin{split}
        \overline{\abs{\mathcal{M}}}_{uu\rightarrow tt}^2 &= \frac{2 (m_t^2-s)^2 - 4 m_t^2 u + 2 u^2}{(m_{Z'}^2-t)^2}+ \\
        & + \frac{2 (m_t^4-2 m_t^2 s-2 m_t^2 t+s^2+t^2)}{(u-m_{Z'}^2)^2} - \\
        &-\frac{4 s (3 m_t^2-s)}{(m_{Z'}^2-t) (u-m_{Z'}^2)}
    \end{split}
\end{equation}
$\overline{\abs{\mathcal{M}}}_{uu\rightarrow tt}^2$ could be expressed in the partonic center-of-mass frame as following:
\begin{equation}
\begin{split}
        \frac{1}{((-2 m_t^2+2 m_{Z'}^2+s^2)^2-s^4 \cos ^2(\theta ))^2} \Bigl(
       &4 s^2 (4 m_t^2+s^2) (-2 m_t^2+2 m_{Z'}^2+s^2)^2+ \\
       &+8 s^4 \cos ^2(\theta ) (2 s^2 (3 m_{Z'}^2-4 m_t^2)+ \\
       &+2 (m_t^2-m_{Z'}^2)^2+7 s^4)+4 s^8 \cos ^4(\theta ) \Bigr)
\end{split}
\end{equation}
where $\theta$ is the scattering angle in the center-of-mass frame.
One can see  for more massive $Z'$, the terms that are proportional to $s$ become more relevant.

\bibliography{sst}
\bibliographystyle{JHEP}

\end{document}